# HYSTERESIS AND AVALANCHES IN THE RANDOM FIELD ISING MODEL

SANJIB SABHAPANDIT

Tata Institute of Fundamental Research
Mumbai 400 005, India

*A thesis submitted to the*

*University of Mumbai*

*for the*

*Ph.D. Degree in Physics*

**2002**

# Acknowledgments

I am deeply indebted to my thesis advisers Deepak Dhar for his constant invaluable guidance and encouragement throughout the last five years. His patience and persistence, insights into various problems, critical thinking and insistence on clarity have been most useful and inspiring. My words will never be adequate to express my gratitude towards him.

Mustansir has always been available for any kind of discussion. I learnt many things from his two courses on statistical mechanics.

Satya has been a very cheerful friend. Discussions with him about physics and non-physics topics have always been very enjoyable.

I have had a nice time with many friends in the Institute. Lots of my free time in TIFR was spent in the enjoyable company of Rajesh, Arun, Dibyendu, Anwesh, Anu, Bhaswati, Ritu, Palit, Patta, Ajay Nandgaonkar. Azizur, Praveen, Vinod, Gulab, Gagan have been good friends since my joining in TIFR. I would like to thank all my other friends, with whom I had numerous weekend (!) parties.

The theory students room has been a very lively place. I would like to thank Goutam, Abhishek, Keshav, Saumen, Justin, Rajesh, Ghosal, Arun, Dibyendu, Nemani, Patta, Kavita, Dariush, Arti, Sumedha, Ramanan, Ajay, Apoorva, Ashik, Ashotosh and Punyabrata for all the enjoyable discussions.

A special word of thanks to Bathija, Girish, Pawar and Shinde for helping out on official matters and for their friendliness.

I acknowledge the "Kanwal Rekhi Scolarship for Career Devlopment" awarded to me by the TIFR Endowment Fund in the academic year 2001-2002.

I would like to thank all the people involved in the development of GNU/Linux and TEX(LATEX), without whom this work would have been very difficult.

Finally, I would like to express my heartfelt gratitude to my parents, and brothers for their love and support.

# Contents

# List of Figures

# Chapter 1

# Introduction

In the recent years, there has been a growing interest in the study of nonequilibrium systems. The states of systems at thermal equilibrium, are a priori given by the Boltzmann-Gibbs distribution i.e. there is an energy function $E(C)$ associated to every possible configuration $C$ of the system and each configuration $C$ has a weight proportional to $\exp[-E(C)/k_B T]$, where $k_B$ is Boltzmann's constant and $T$ is the temperature of the system. Then all the thermodynamic quantities are obtain by averaging over all the configurations with respective weights. However, in nature there exist a wide variety of systems, which are not in thermal equilibrium. The probabilities of different states of these systems are not given by the Gibbs distribution, but are determined by the underlying microscopic dynamical processes, and are often hard to determine due to lack of a general framework.

An important class consists of nonequilibrium systems, which when driven by slowly varying external forcing, relax through avalanche-like dynamics in response to the external perturbations. Examples include sand or rice piles, forest fires, earthquakes, vortices in dirty type II superconductors, solid on solid friction, moving of interfaces in random media, disordered ferromagnets and many others (see Jensen 1998, chap. 3). Depending on the system, the avalanche is characterized by different physical quantities. For example, in sand piles the system is driven by slowly adding sand grains to the system and the avalanche is characterized by the number of sand grains displaced after adding a single grain or the lifetime of the avalanche. In earthquakes, it is the energy release and in case of ferromagnets it is the size of the domain that flips. The avalanches occurs in various sizes in a random sequence, and one is generally interested in the distribution of the avalanche sizes.

What is common in all the systems mentioned above is the existence of threshold and multiple metastable states i.e. if the applied external force is less than a critical value the system does not response and when the force exceeds the critical value the system passes from one metastable state to another. Due to existence of multiple metastable states, the state of a system at a given time depends on the history of evolution (path along which the system is evolving in configuration space) and systems exhibit hysteresis in the zero frequency limit of external forcing[†].

In this thesis, we study a spin model in the presence of disorder, called random field Ising model, introduced by Sethna et al. (1993) in the context of Barkhausen noise and hysteresis in disordered ferromagnets. In this model, as the external field increases, the magnetization increases as groups of spins flip up together. The dynamics is governed by the existence of many metastable states, with large energy barriers separating different metastable states. We hope that this study of non-equilibrium response in this model would help in the more general problem of understanding the statistical mechanics of metastable states in glassy systems.

The remainder of this chapter is organized as follows. section 1.1 contains a brief review of theoretical studies of hysteresis in ferromagnets. In section 1.2, we briefly discuss Barkhausen effect. In section 1.3, we define the random field Ising model with zero temperature dynamics, and discuss some earlier results. In section 1.4, we discuss some of the equilibrium properties of random field Ising model. Section 1.5 gives an outline of the remaining chapters.

## 1.1  Hysteresis in ferromagnets

The studies of hysteresis in magnetic materials has been there in various branches of science, for a long time (see Bertotti 1998). Apart from the intellectual interest, it also has wide range of technological applications, from designing transformer cores to memory devices.

Physicists have been looking for a convincing general theory to interpret the phenomenon of hysteresis in magnets since the time of Rayleigh (1887), who gave the first phenomenological theory where the experimental magnetization curves at small field were approximated by parabolas. Starting from the demagnetization state (zero magnetization in the absence of external field), the magnetizations $M_\pm$ at small fields $\pm h$, are expressed

---

[†]Systems also show hysteresis under periodic forcing. For example, when a ferromagnet is placed in oscillation field, the magnetization lags behind its instantaneous equilibrium value and gives rise to hysteresis loop. But the area of the hysteresis loop tends to zero in the zero frequency limit of the driving field (Dhar and Thomas 1992)

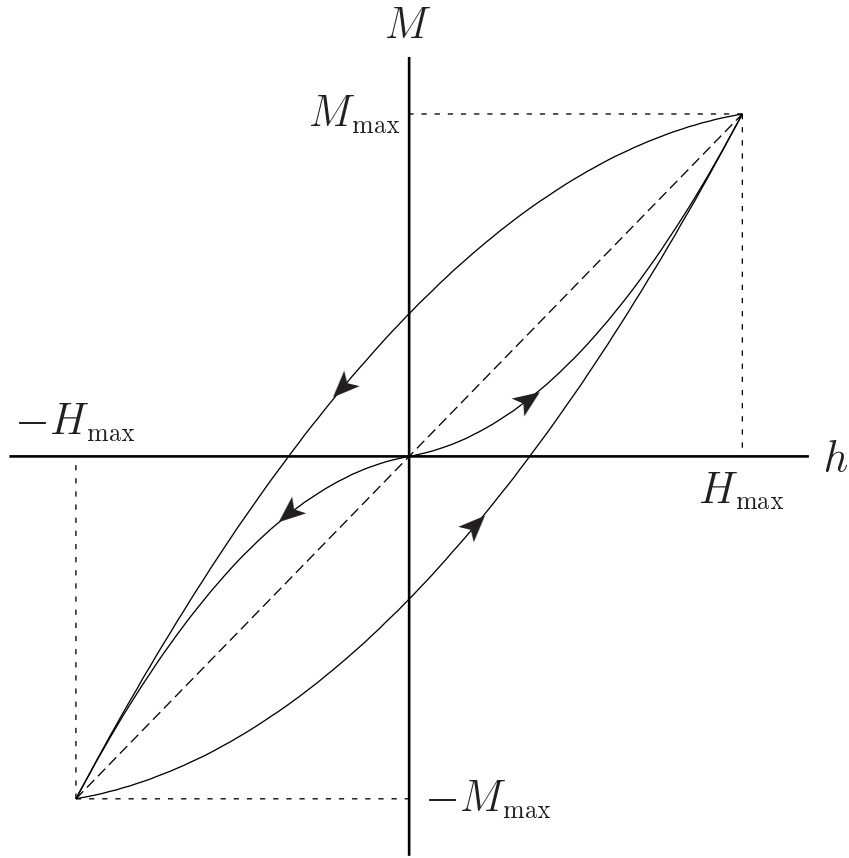

**Figure 1.1:** Rayleigh hysteresis loop

as
$$M_{\pm} = ah \pm bh^2, \tag{1.1a}$$

and when the field is cycled between small $\pm H_{\text{max}}$, the lower and the upper curves of the hysteresis loop are represented by

$$M_{\pm} = (a + bH_{\text{max}})h \pm \frac{1}{2}b(h^2 - H_{\text{max}}^2). \tag{1.1b}$$

In Fig. 1.1, we have shown the magnetization curves starting from the demagnetization state given by Eq. (1.1a) and the Rayleigh hysteresis loop given by Eq. (1.1b).

In Weiss's (1907) theory of ferromagnetism, he postulated the existence of a powerful internal "molecular field" in ferromagnet materials, which would tend to tries to align the magnetic moments along one direction. It agrees with some of the experimental cases where it is possible to attain a large saturation magnetization by the application of a very weak magnetic field [see Fig. 1.2]. However, it did not explain the fact that, it is also possible for the magnetization to be zero (or nearly zero) in the absence of a magnetic field.

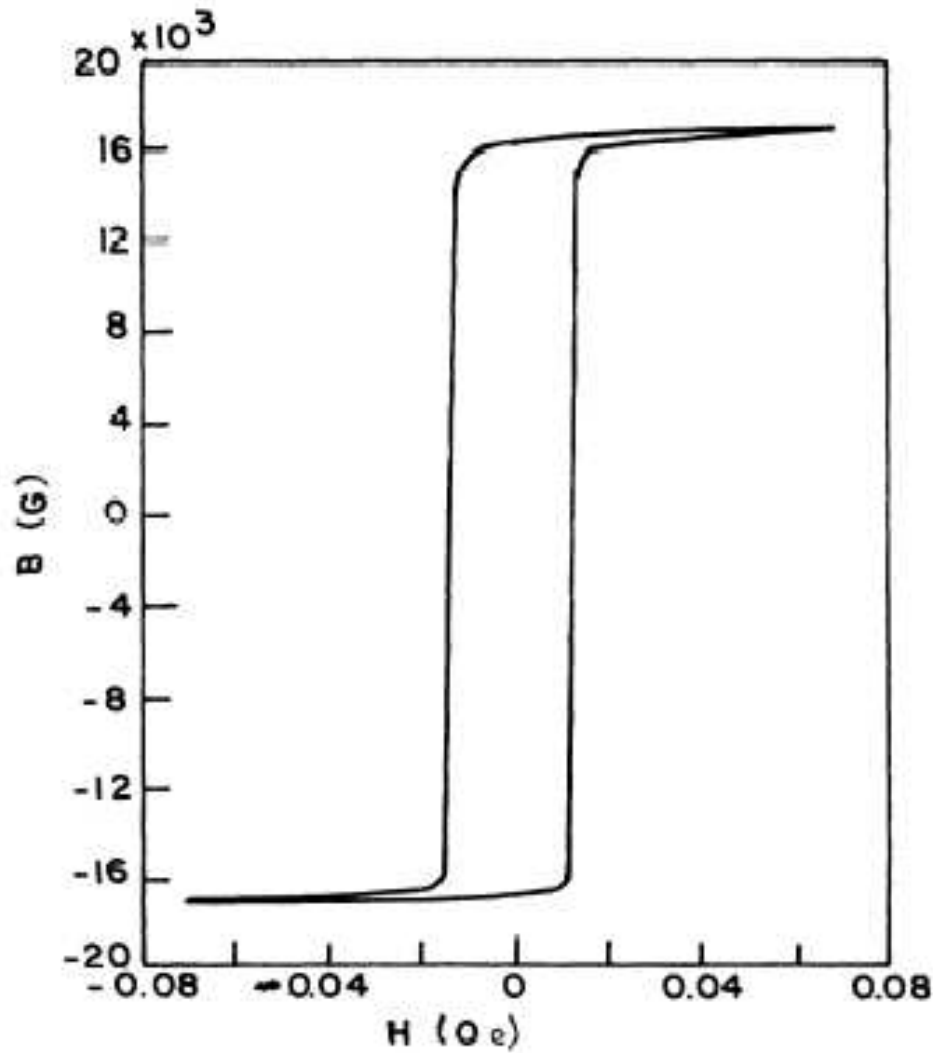

**Figure 1.2:** Magnetization curve of single crystal of silicon iron. (Williams and Shockley 1949).

Weiss further made an assumption that, a ferromagnet can be subdivided into regions called *magnetic domains*. In each individual domain, the the magnetic moments are aligned along the molecular field, but the orientation of the spontaneous magnetizations in each domain distributed randomly inside the sample and hence the resultant magnetization could be zero in the absence of a external field, even at very low temperature. If a external field is applied opposite to the magnetization direction, a domain reverses the direction of magnetization when the external field exceeds a critical value $H_c$. Therefore, if a gradually increasing external field is applied, domains whose magnetization vectors are at an angle $(\pi - \theta)$ with the external field, will suddenly reverse direction when the external field exceeds $H_c/\cos\theta$. This results a finite bulk magnetization for external field $h > H_c$. A comprehensive review of the physical principles of domain theory and of the crucial experiments which bear directly on the foundation of the subject, may be found in an article by Kittel (1949).

Preisach (1935) introduced a modified domain model, in which he assumed that material is composed of many small domains and each of them possesses a rectangular hysteresis loop. The interaction between domains are represented by a local field acting on each domain. Thus each domain has two different coercive fields $\alpha$ and $\beta$ for the increasing and decreasing branches respectively. The ensemble of domains is then described by the distribution function $P(\alpha, \beta)$ of the values of $\alpha$ and $\beta$ and hysteresis loops are obtained by taking the weighted sum of magnetization in all the domains.

Sethna et al. (1993) proposed the random field Ising model with the zero temperature dynamics as a simple theoretical model for the Barkhausen noise and hysteresis in disordered ferromagnets. In this model, magnetic domains are represented by Ising spins ($s_i = \pm 1$) and the external field is coupled to these spin. In contrast to the Preisach model of hysteresis, where interactions between the individual hysteresis units (domains) are ignored, in the random field Ising model the spins interact ferromagnetically with their neighbors. The homogeneities and disorder in materials are modeled by introducing a uncorrelated random field acting on each domain, chosen at random from some distribution. Since the domains interact ferromagnetically, flipping of a domain at some external field may force the neighboring domains to flip as well in the same direction, thus leading to an avalanche of domain flips, which is analogue of a Barkhausen pulse in real magnets (for a comparison, see Fig. 1.5 and Fig. 1.6).

In the models discussed above, the hysteresis does not depends on the rate at which the external field is varied i.e. relaxation time from one metastable state to another is much larger to the rate at which the system is driving. In contrast, there are also other models studied in the context of rate-dependent hysteresis, where the system exhibits hysteresis only when it is driven at a finite rate (see Chakrabarti and Acharyya 1999, for a recent review). Hysteresis in the $N$-vector model was widely studied by many authors (Rao et al. 1990, Dhar and Thomas 1992, 1993, Somoza and Desai 1993). It was shown that in all dimensions $d > 2$, for $N \geq 2$ at low frequency $\omega$ and low amplitude $H_0$ of the driving field the area of the hysteresis loop scales as $(H_0\omega)^{1/2}$ with logarithmic corrections. At high frequencies the area varies as $H_0^2/\omega$. For any $H_0$, there is a dynamical phase transition separating these two frequency regimes. Above the critical frequency $\omega(H_0)$, the hysteresis loop does not posses inversion symmetry. Using the nucleation theory Dhar and Thomas (1993) showed that for $N = 1$ and $d > 1$, the area of the hysteresis loop scales as $|T \ln(H_0\omega)|^{-1/(d-1)}$ for $\omega \ll H_0$.

## 1.2 Barkhausen effect

The first evidence for the existence of ferromagnetic domains was from an experiment by Barkhausen (1919). His experiment consists in amplifying the voltage induced in a secondary pick-up coil wound around a ferromagnetic sample, while the sample is being magnetized by a continuous variation of external magnetic field [Fig. 1.3]. He observed a noise induced in the pick-up coil, corresponds to a sudden, discontinuous jumps in magnetization [Fig. 1.4]. These jumps are interpreted as discrete changes in the size or rotation of ferromagnetic domains. An elementary introduction of the Barkhausen effect may be found in the textbook by Feynman et al. (1977).

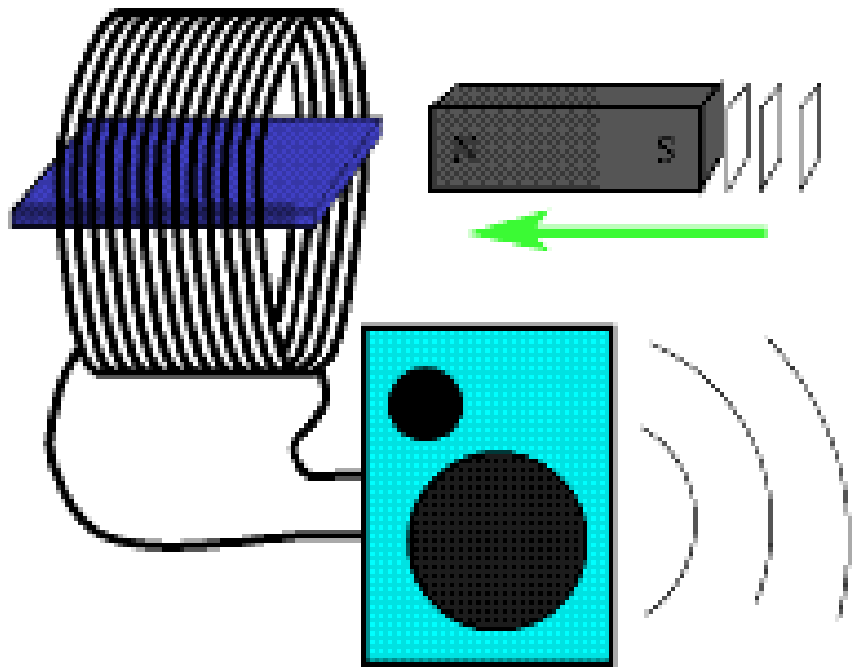

**Figure 1.3:** Barkhausen effect

In the recent years, there has been a great interest in the study of the statistical properties of the Barkhausen noise. A typical train of barkhausen noise signals observed in experiments is shown in Fig. 1.5. Three basic physical quantities that describe a single Barkhausen noise signal in an experiment, are signal duration, area of the signal and the energy released during the signal occurrence. It is observed that distribution of these quantities follow power law over a few decades with a cut-off as shown in Fig. 1.7 (see Spasojević et al. 1996, and references therein). This power law tail in the Barkhausen avalanche distribution was interpreted by Cote and Meisel (1991) as an example of *self-organized*

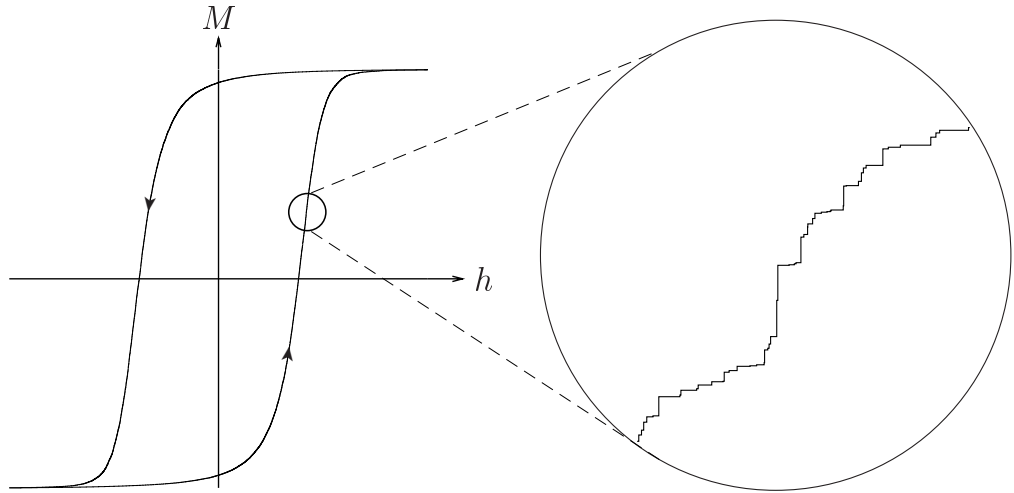

**Figure 1.4:** Barkhausen jumps

*criticality*[†]. But Perković et al. (1995) have argued that large bursts are exponentially rare, and the approximate power-law tail of the observed distribution comes from crossover effects due to nearness of a critical point.

Barkhausen effect is also widely used as a noninvasive material characterization technique for ferromagnetic materials (see Sipahi 1994, for an overview).

## 1.3  Hysteresis in random field Ising model

The nonequilibrium random field Ising model was proposed by Sethna et al. (1993) as a model for Barkhausen noise and hysteresis in ferromagnets. The model is defined on a lattice. At each lattice site $i$, there is a Ising spin $s_i = \pm 1$, which interacts with nearest neighbors through a ferromagnetic exchange interaction ($J > 0$). Spins $\{s_i\}$ are coupled to the on-site quenched random magnetic field $h_i$ and the external field $h$. The Hamiltonian of the system is given by

$$H = -J \sum_{\langle i,j \rangle} s_i s_j - \sum_i h_i s_i - h \sum_i s_i, \quad (1.2)$$

where $\langle i, j \rangle$ denotes that the sum runs over nearest neighbor pairs of spins on sites $i$ and $j$. We assume that $\{h_i\}$ are quenched independent identically distributed random variables with the probability that the value of the random field at site $i$ lies between $h_i$ and $h_i + dh_i$ being $\phi(h_i)\, dh_i$.

---

[†]In *self-organized critically*, systems exhibit critical behavior (power law correlations), without fine tuning any parameter (for an overview, see: Dhar 1999, Jensen 1998, Bak 1997).

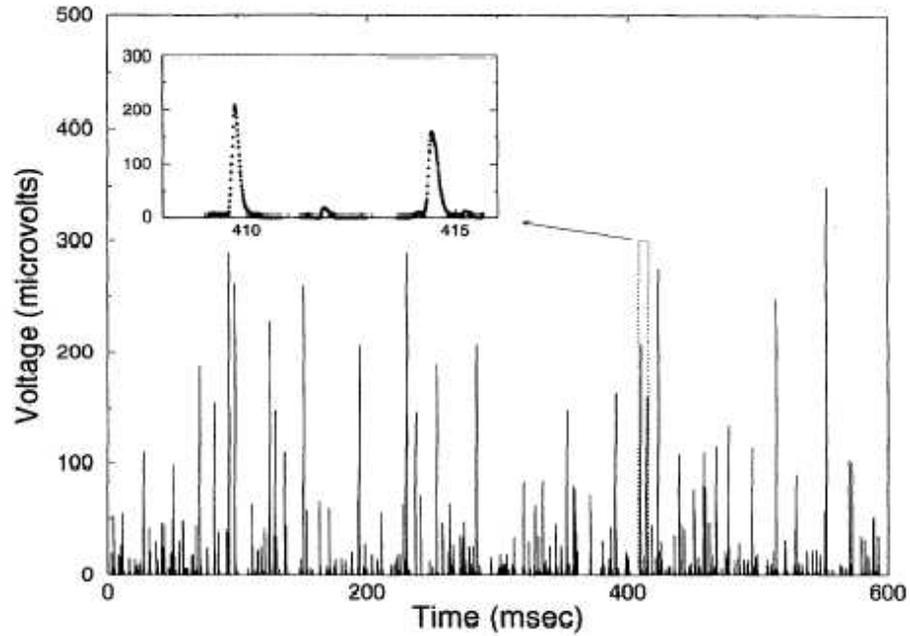

**Figure 1.5:** An example of experimental Barkhausen signal (voltage pulse produce from a pickup coil around a ferromagnet subjected to a slowly varying applied field) (Urbach et al. 1995).

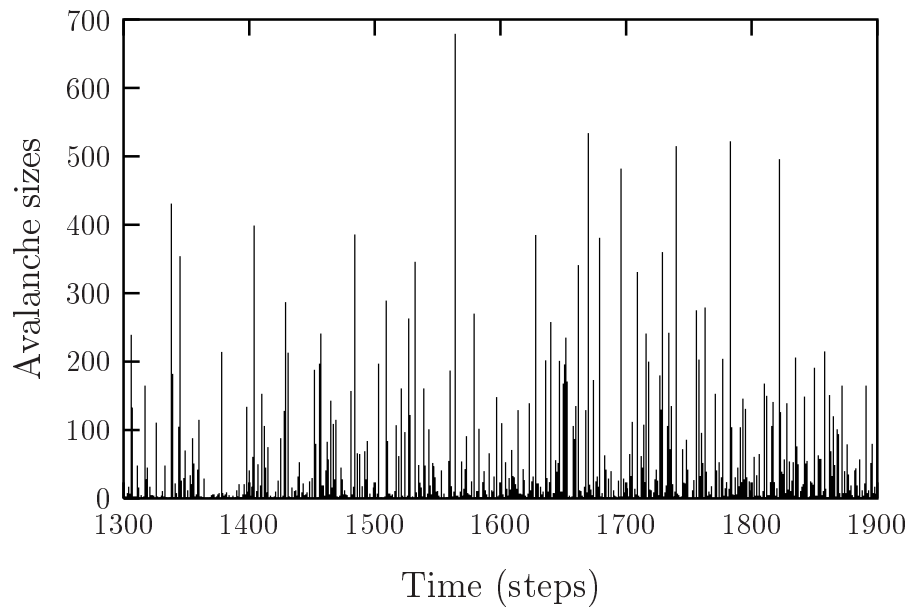

**Figure 1.6:** Time series of the avalanches (the number of spin flips at a given field) in the random field Ising model on a square lattice of size $200 \times 200$. From one avalanche to the next avalanche is considered as one time step.

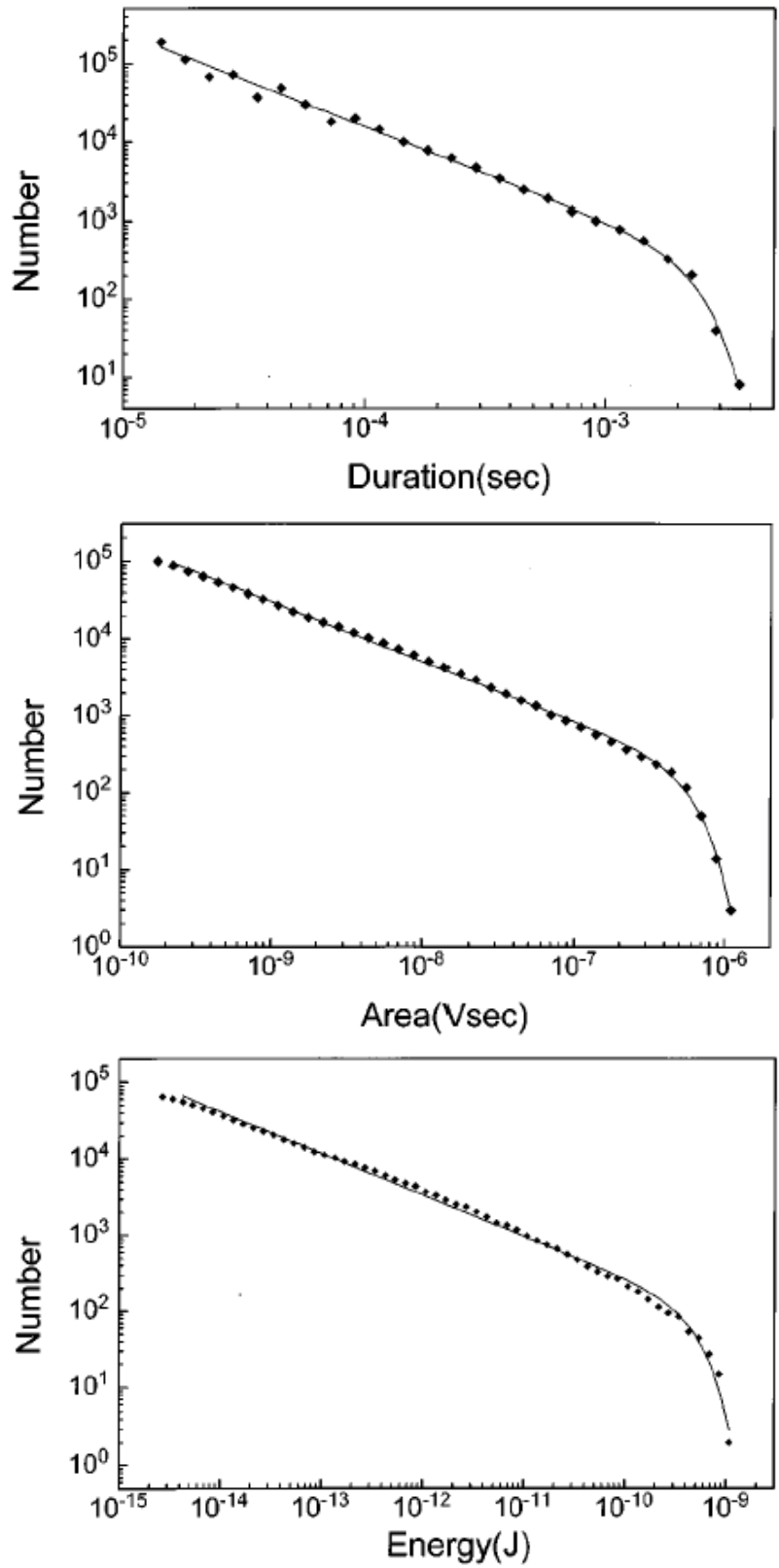

**Figure 1.7:** Experimental data for the distribution of Barkhausen signal durations, areas and energies (Spasojević et al. 1996).

When the external field is changed, the system relaxes to a stable spin configuration through zero-temperature Glauber single-spin-flip dynamics (Kawasaki 1972, see), which is specified by the transition rates

$$\text{Rate}[s_i \to -s_i] = \begin{cases} \Gamma & \text{if } \Delta E \leq 0 \\ 0 & \text{otherwise} \end{cases} \quad (1.3)$$

where $\Delta E$ is the change of energy in the system as a result of the spin flip. Therefore, a spin-flip is allowed only if the process lowers energy. We assume that the external field is increased adiabatically, i.e. $\Gamma \gg \omega$, the rate at which the magnetic field $h$ is increased. Thus if the spin-flip is allowed, it relax instantly, so that the spin $s_i$ in a stable configuration is parallel to the net local field $\ell_i$ at the site:

$$s_i = \text{sign}(\ell_i) = \text{sign}\left(J\sum_{j=1}^{z} s_j + h_i + h\right). \quad (1.4)$$

Note that the limit $\omega/\Gamma \to 0$ is taken after the limit $T \to 0$. If the limits are taken in the reverse order, the state of the the system at each $h$, is the equilibrium state for all finite $T$ and the hysteresis loop area goes to zero.

We start with $h = -\infty$, when all spins are down and slowly increase $h$. As we increase $h$, some sites where the quenched random field is large positive will find the net local field positive, and the spin at that site will flip up. Flipping a spin makes the local field at neighboring sites increase, and in turn may cause them to flip. Thus, the spins flip in clusters of variable sizes. If increasing $h$ by a very small amount causes $s$ spins to flip up together, we shall call this event an avalanche of size $s$. As the applied field increases, more and more spins flip up until eventually all spins are up, and further increase in $h$ has no effect.

As an illustrative example, consider a four by four square lattice with periodic boundary condition and a particular realization of quenched random fields, which is shown in Fig. 1.8(a). We set $J = 1$. Now start with $h = -\infty$, when all spins are down [Fig. 1.8(b)] and slowly increase it. A spin with $m$ up neighbors, flips up at $h$, if the quenched random field at the particular site $h_i > 4 - 2m - h$. Therefore, when the external field just exceeds the value $1.1$, the local field at the site where $h_i = 2.9$, becomes positive and the spin at that site flips up. This increases the local fields at its neighboring sites by $2J$ and as a result spins at some of these sites flip up [Fig. 1.8(c)]. These process continues till there is no more sites where the local field is positive at that external field. In the figure, we denote active spins with different color according to order at which they flip during the avalanche. As shown in Fig. 1.8(b) - (f), the system passes from one stable configuration to another, as the external field is increased, till all the spins in the system flip up. In Fig. 1.8(g), we show the different clusters of spins, which flip at different fields. Fig. 1.8(i) shows the corresponding magnetization.

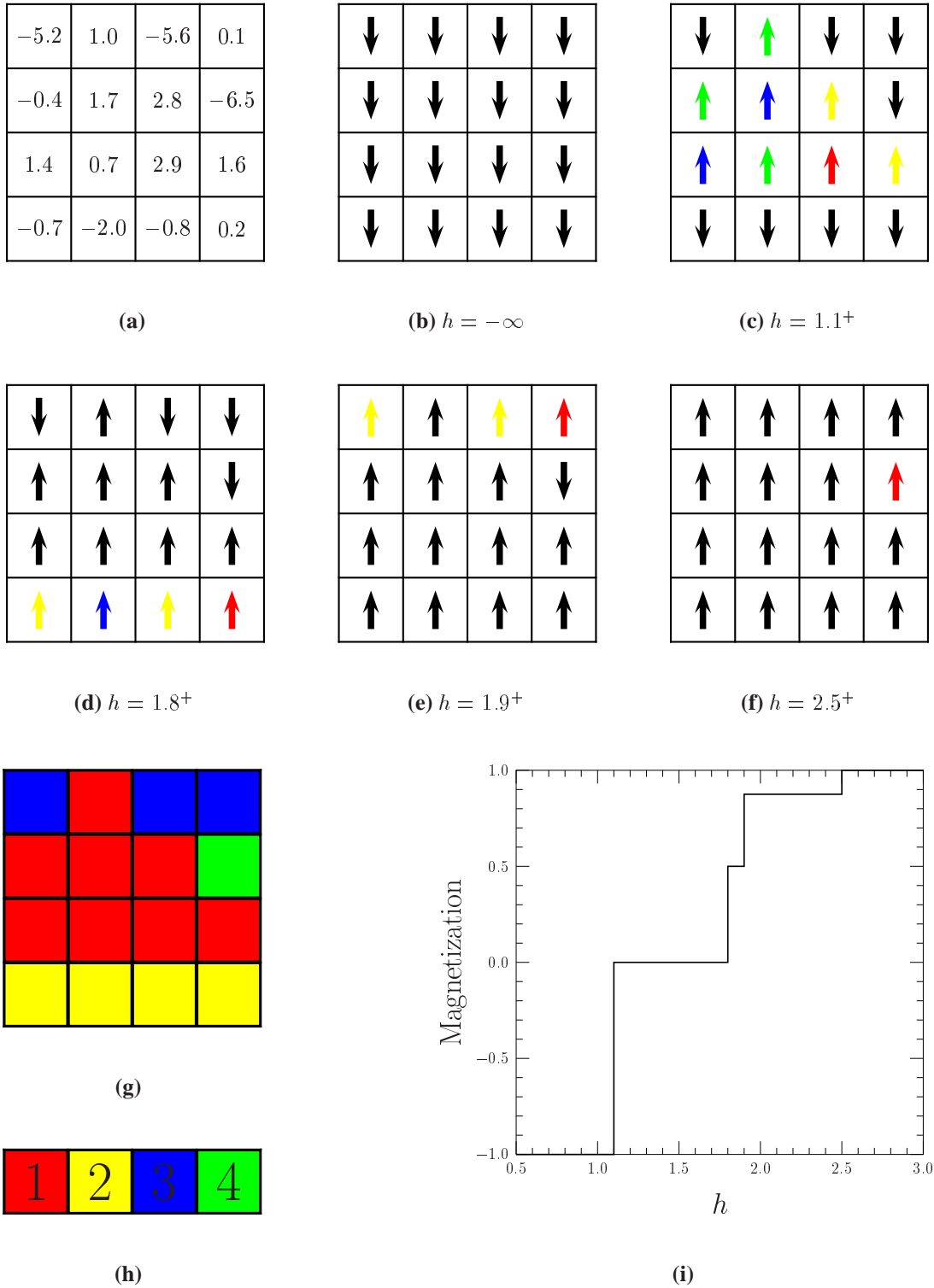

**Figure 1.8:** (a) Quenched random fields at various sites. (b) - (f) Stable spin configurations at various external fields. The black spins are inactive spins at that particular field and the colored spins are part of the avalanche. The colors specify the order in which spins flip during the avalanche. (g) Clusters of spins, which flip during one avalanche. (h) Color map showing the order of events. (i) Magnetization curve, corresponding to evolution.

Sethna et al. (1993) studied the model with the infinite-range interaction (mean field theory), where every spin is coupled to all $N$ other spins with coupling $J/N$. They found that there exists a critical value of disorder $\Delta_c$ (which in the case of a Gaussian distribution of random fields is $= \sqrt{(2/\pi)}J$ ), below which the hysteresis curve displays a jump due to an infinite avalanche of spin flips, which spans the system. Above the critical disorder systems show smooth magnetization curve without macroscopic jumps. However, this mean field theory does not show any hysteresis for disorder $\Delta \geq \Delta_c$. Dahmen and Sethna (1993, 1996) studied the hysteresis loop critical exponents expanding about mean field theory in $6-\epsilon$ dimensions. A power-low distribution with avalanche of all sizes is seen only at the critical value of the disorder. However, the numerical simulations by Perković et al. (1995) indicate that the critical region is remarkably large: almost three decades of power-law scaling in the avalanche size distribution remain when measured 40% away from the critical point. Therefore, they argued that several decades of scaling seen in experiments need not be self-organized criticality, as many of the samples might have disorders within 40% of the critical value.

Interestingly the model can be solved exactly on a Bethe lattice for the magnetization on the hysteresis curve for arbitrary distribution of random fields (Dhar et al. 1997). In contrast to the infinite-ranged mean field theory, the calculation on Bethe lattice shows hysteresis even for large disorder. Another interesting result of the Bethe lattice calculation is that, the first order jump in the magnetization disappear for coordination number of the lattice less than 4. Only for coordination number 4 and above, there exists a critical value of disorder below which there is a jump discontinuity in the magnetization.

## 1.4 Equilibrium properties of random field Ising model

In this thesis we are interested in the nonequilibrium properties of the random field Ising model. However, it is useful to recall the equilibrium properties of this model, which has been an important problem in statistical physics for a long time. This model has a number of interesting realization in nature. A recent review of earlier work on this model may be found in an article by Nattermann (1998). This model was first studied by Imry and Ma (1975), in the context of possible destruction of long-range order by arbitrarily weak quenched disorder. The pure Ising model with nearest neighbor interaction $H_0 = -J \sum_{\langle i,j \rangle} s_i s_j$, is known to have a ferromagnetically ordered phase in all dimensions $d > 1$. When the random field term $-\sum h_i s_i$ is introduced, it acts against the order. Imry and Ma (1975) argued that arbitrarily weak disorder destroys long-ranged ferromagnetic order in dimensions $d \leq 2$. The argument goes as follows:

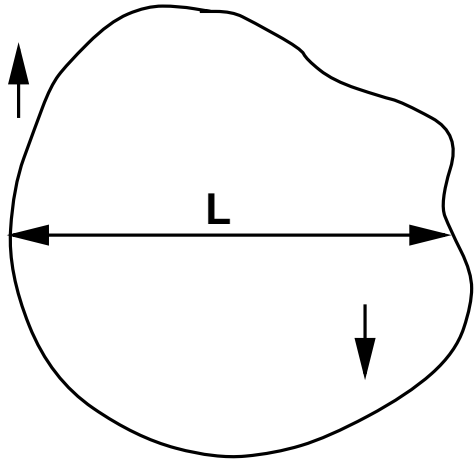

**Figure 1.9:** Domain of reverse spins.

If we consider a domain of reverse spins [Fig. 1.9], of linear size $\sim L$, the domain wall energy is $\sim L^{d-1}$. However, according to the central limit theorem, if the random field has short-range spatial correlations, the fluctuation in the magnetic field energy in such domains is $\sim L^{d/2}$. Thus, by splitting into domains of size $L$, the system will gain bulk energy of $\mathcal{O}(L^{d/2})$ per domain, and loose a surface energy a surface energy of $\mathcal{O}(L^{d-1})$ per domain. Thus, whenever $d \leq 2$, there will exists a large enough $L$, for an arbitrarily small random field, where it will become energetically favorable to the system to break into domains of that size.

The argument by Imry and Ma (1975) suggests that the *lower critical dimension*[†] is $d_l \geq 2$, rather than $d_l = 2$, because other mechanisms could destroy long-range order in higher dimensions. It is widely believed that the *upper critical dimension*[‡] is $d_u = 6$, instead of $d_u = 4$ for the pure Ising system. However, whether the lower critical dimension $d_l = 2$ or $d_l = 3$, was a matter of a long controversy, but has now been established that $d_l = 2$. Imbrie (1984) showed that if the disorder is small, the model in dimension $d = 3$ exhibits long-range order at zero temperature. Aizenmann and Wehr (1989) rigorously proved uniqueness of the Gibbs state in $d = 2$, i.e. absence of any phase transition, in agreement with the Imry-Ma prediction.

---
[†]The dimension below which long-range ferromagnetic order cannot exist.
[‡]The dimension above which the critical exponents are those of the Gaussian fixed point.

As far as an exact calculation of thermodynamic quantities is concerned, there are only a few results. For example, Bruinsma (1984) studied the random field Ising model on a Bethe lattice in the absence of an external field and for a bivariate random field distribution. There are no known exact results for the average free energy or magnetization, for a continuous distribution of random field, even at zero temperature and in zero applied field.

## 1.5  Outline of this thesis

In this thesis, we study the nonequilibrium ferromagnetic random field Ising model with zero temperature Glauber single flip dynamics. The remaining chapters of the thesis are organized as follows:

In chapter 2, we discuss some special properties of the model that makes the analytical treatments possible. We briefly recapitulate the derivation of self-consistent equations for the magnetization in the model.

In chapter 3, we use a similar argument to construct the generating function for the avalanche distribution for arbitrary distribution of the quenched random field. In section 3.2, we consider the special case of a rectangular distribution of the random field. In this case, we explicitly calculate the probability distribution of avalanches, for the for Bethe lattices with coordination numbers $z = 2$ and $3$. In section 3.3, we analyse the self-consistent equations to determine the form of the avalanche distribution for some general unimodal continuous distributions of the random field. In chapter 4, we derive the self-consistent equations for the magnetization on minor hysteresis loops on a Bethe lattice, when the external field is varying cyclically with decreasing magnitudes. We also discuss some properties of stable configurations, when the external field is varying.

In chapter 5, we study the model with an asymmetric distribution of quenched fields, in the limit of low disorder in two and three dimensions. We relate the spin flip process to bootstrap percolation, and find nontrivial dependence of the coercive field on the coordination number of the lattice.

Chapter 6 contains a discussion of our results, and some concluding remarks.

Some algebraic details of the analytical solution for the distribution of avalanche sizes, for the rectangular distribution of quenched fields on Bethe lattice are relegated to appendix A.

Part of this work has appeared in journals as refereed papers. Though it is mostly a repetition of the material presented in chapter 3 and chapter 5, for the convenience of the reader, we have reproduced these papers as an appendix (reprints) to this thesis.

# Chapter 2

# Earlier exact results on hysteresis in random field Ising model

The difficulty of solving various mathematical equations describing actual physical situations leads to various approximation methods. These approximation method can be classified into two categories: One in which the approximation is made in the mathematical equations itself and another in which the physical system is simplified. Into the second category fall many lattice model systems. Again in higher dimensions the lattices contain closed circuits which makes the model difficult to solve. Thus, one considers the problem in the mean-field theory, where the underlying lattice structure becomes irrelevant or on a different lattice where it can be solved exactly. Bethe lattice or Cayley tree is one in which there is no circuits at all which makes the model easier to solve. The simplicity of the lattice motivates one to study various systems on a Bethe lattice.

In this chapter we briefly discuss derivation of hysteresis curve in the random field Ising model in the mean field theory (infinite-range interaction) and on the Bethe lattice. In sections 2.1 and 2.2 we discuss two properties of zero temperature random field Ising model, namely the *return point memory effect* and the *abelian nature of spin-flips*, which is used to set up self-consistent equation to determine magnetization on the Bethe lattice and later in other chapters.

## 2.1 Return point memory

Sethna et al. (1993) showed that the RFIM exhibits the following return point memory effect: Suppose we start with $h = -\infty$, and all spins down at $t = 0$. Now we change the field slowly with time, in such a way that $h(t) \leq h(\mathcal{T})$, for all times $t < \mathcal{T}$. Then the configuration of spins at the final instant $t = \mathcal{T}$ does not depend on the detailed time dependence of $h(t)$, and is the same for all histories, so long as the condition $h(t) \leq h(\mathcal{T})$ for all earlier times is obeyed. In particular, if the maximum value $h(\mathcal{T})$ of the field was reached at an earlier time $t_1$, then the configuration (and hence the magnetization) at time $\mathcal{T}$ is exactly the same as that at time $t_1$ [Fig. 2.1].

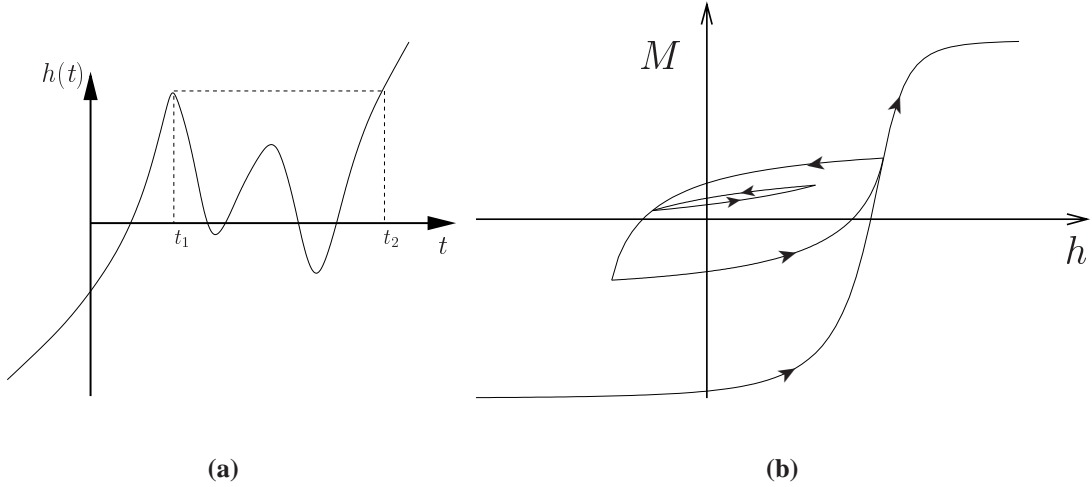

**Figure 2.1:** (a) change of external field with time. (b) magnetization curve vs. external field given by (a). When the external field returns to the previous extremal value, the magnetization returns to the value at that field i.e. $M(h(t_1)) = M(h(t_2))$.

Consider two spin configurations $C\{s_1, s_2, \ldots, s_n\}$ and $C'\{s'_1, s'_2, \ldots, s'_n\}$. If $s_i \geq s'_i$ for each site $i$, the configurations $C$ and $C'$ are called partially ordered, $C \geq C'$. Let two configurations $C(t)$ and $C'(t)$ be evolve under the field $h(t)$ and $h'(t)$ respectively. Suppose the initial configurations $C(0)$ and $C'(0)$ are partially ordered such that $C(0) \geq C'(0)$ and the fields satisfy $h(t) \geq h'(t)$. Then if a spin $s'_i(t)$ is up in configuration $C'(t)$, the corresponding spin $s_i(t)$ in configuration $C(t)$ must be up, since the local field $\ell_i(t)$ in $C(t)$ can not be less than $\ell'_i(t)$ in $C'(t)$. Therefore the configurations $C(t)$ and $C'(t)$ will always remain partially ordered, $C(t) \geq C'(t)$. This is the no passing property of the system. An earlier treatment of "no passing" rule was given by Middleton (1992) in the context of charged-density waves.

Let us consider the Fig. 2.2. The configuration $A$ is reached by increasing the field from a lower value to $h_1$. On increasing the field monotonically from $h_1$ to $h_2$, configuration $B$ is reached. Naturally, the configurations $A$ and $B$ are partially ordered such that

$$B \geq A. \tag{2.1a}$$

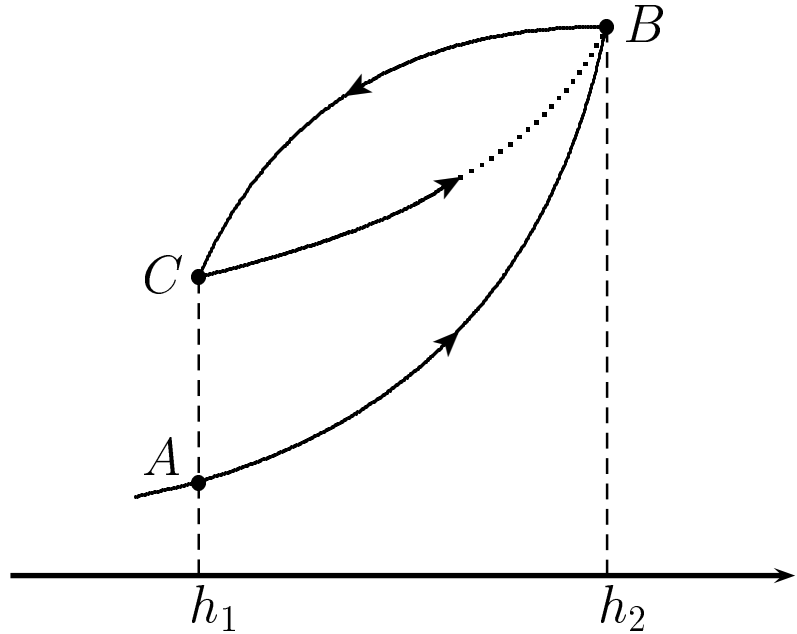

**Figure 2.2:** Partial ordering between configurations, when the field is (1) increased from $h_1$ to $h_2$, (2) then decreased from $h_2$ to $h_1$, (3) again increased from $h_1$ to $h_2$.

Similarly when the field is decreased monotonically from $h_2$ to $h_1$, partial ordering exists between the configuration $B$ and the final configuration $C$ such that

$$C \leq B. \tag{2.1b}$$

Since during the evolution from $A$ to $C$, the field $h(t)$ satisfies $h(t) \geq h_1$,

$$C \geq A. \tag{2.1c}$$

Now suppose the configuration $C$ evolve to a configuration $D$ when the field is again increased monotonically from $h_1$ to $h_2$. Since the partial ordering is preserved by dynamics, from Eq. (2.1b),

$$D \leq B, \tag{2.1d}$$

and from Eq. (2.1c),

$$D \geq B, \tag{2.1e}$$

as $A$ evolves to $B$, when the field is increased from $h_1$ to $h_2$. From Eq. (2.1d) and Eq. (2.1e), we must have

$$D = B, \tag{2.1f}$$

i.e. the system returns exactly to the same earlier configuration, when the field is decreased from a maximum value $h_2$ and then increased to the same value. The same memory effect extends to subcycles within the cycles and so on.

## 2.2 Abelian property

Because of the previous property, we may choose to increase the field suddenly from $-\infty$ to $h(T)$ in a single step. Then, once the field becomes $h = h(T)$, several spins would have positive local fields. Suppose there are two or more such flippable sites. Then flipping any one of them up can only increase the local field at other unstable sites, as all couplings are ferromagnetic. Thus to reach a stable configuration, all such spins have to be flipped, and *the final stable configuration reached is the same, and independent of the order in which various spins are relaxed.* This is the *abelian property* of relaxation (Dhar et al. 1997). Using the symmetry between up and down spins, it is easy to see that the abelian property also holds whether the new value of field $h''$ is greater or less than its initial value $h'$ so long as one considers transition from a stable configuration at $h'$ to a stable configuration at $h''$.

## 2.3 Hysteresis in the infinite-range interaction model

In this section, we will briefly discuss the results obtained by Sethna et al. (1993), on the hysteresis in the random field Ising model with infinite-range interaction. In this mean field theory, every spin is coupled to all $N$ other spins with coupling $J/N$. The Hamiltonian is given by

$$H = -\frac{J}{N}\left(\sum_i s_i\right)^2 - h\sum_i s_i - \sum_i h_i s_i. \tag{2.2a}$$

Now the interaction of a spin with other spins is replaced by its interaction with the magnetization $M(h)$ of the system. The Hamiltonian then takes the form

$$H = -\sum_i (JM + h + h_i)s_i, \tag{2.2b}$$

i.e. the effective local field at site is $JM + h + h_i$. The spin at this site will flip up if this field is positive i.e. the quenched random field $h_i$ at this site exceeds $-JM - h$. This

happens with probability
$$\int_{-JM(h)-h}^{\infty} \phi(h_i)\,dh_i.$$
Therefore, the average magnetization satisfies the self-consistent equation
$$M(h) = 2\int_{-JM(h)-h}^{\infty} \phi(h_i)\,dh_i - 1. \tag{2.3}$$
Note that, for symmetric distributions of random fields, $M(0) = 0$ is the trivial solution at $h = 0$. Now, if $M(0) = 0$ is the only solution at $h = 0$, then there is no hysteresis. To have other nontrivial solutions for $M(0)$, the slope of the expression on the right hand side of Eq. (2.3), as a function of $M(0)$ must be greater than unity at $M(0) = 0$. At $h = 0$ and near $M(0) = 0$, the right hand side of Eq. (2.3) can be approximated as $2M(0)J\phi(0)$. Therefore, the condition that the Eq. (2.3) to has multi-valued solution is
$$\phi(0) \geq \frac{1}{2J}. \tag{2.4}$$
This condition corresponds to a critical disorder strength $\Delta_c$ (width of the random field distribution), above which there is no hysteresis i.e. the magnetization follows the same curve in the increasing and decreasing field. Below $\Delta_c$, the magnetization curves in the increasing and decreasing field are different near $h = 0$, i.e. the system exhibits hysteresis. Moreover, there is a critical field, where the magnetization jumps from one solution to another one. In a specific case, where the random field distribution is Gaussian,
$$\phi(h_i) = \frac{1}{\sqrt{2\pi}\Delta} \exp\left(-\frac{h_i^2}{2\Delta^2}\right), \tag{2.5}$$
using the condition given by Eq. (2.4), the critical value of disorder is obtained as
$$\Delta_c = \sqrt{(2/\pi)}J. \tag{2.6}$$
Figure 2.3 shows the magnetization curves for this mean-field at various values of disorder $\Delta < \Delta_c$, $\Delta = \Delta_c$ and $\Delta > \Delta_c$ for Gaussian distribution of random fields. Note that hysteresis and jump in the magnetization exist only below a critical disorder [Fig. 2.5(c)].

Sethna et al. have studied in detail the case of critical disorder, and the power-law divergence of various quantities at this critical point. The special value of disorder does not seem to be particularly important and we shall not discuss it here.

## 2.4  Hysteresis on the Bethe lattice

The shortcoming of the treatment discussed in the previous section is that the pair couplings are weak and no correlations and short-range order. One can keep mean field theory, but add correlations by working on a Bethe Lattice.

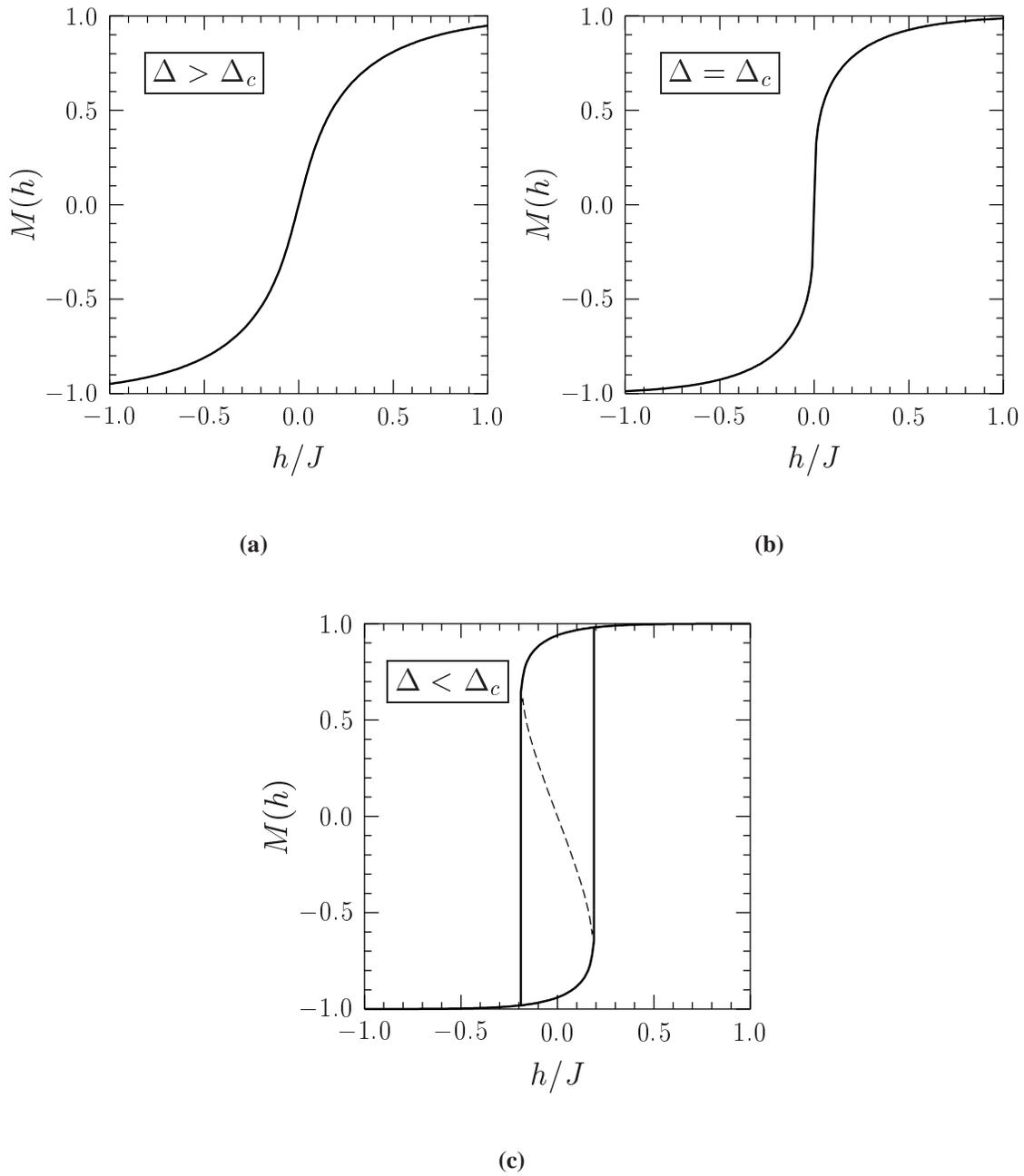

**Figure 2.3:** Magnetization curve for the random field Ising model with infinite-range interaction at various values of disorder: (a) $\Delta = J$, (b) $\Delta = \Delta_c = \sqrt{(2/\pi)}J$ and (c)$\Delta = 0.5J$, for the Gaussian random field distribution given by Eq. (2.5). The dashed line in (c) shows the third root of the self-consistent equation for magnetization, given by Eq. (2.3).

The advantage of working on the Bethe lattice is that the usual BBGKY hierarchy of equations for correlation functions closes, and one can hope to set up exact self-consistent equations for the correlation functions. The fact that Bethe's self-consistent approximation becomes exact on the Bethe lattice is useful as it ensures that the approximation will not violate any general theorems, e.g. the convexity of thermodynamic functions, sum rules etc. In the presence of disorder, in spite of the closure of the BBGKY hierarchy, the Bethe approximation is still very difficult, as the self-consistent equations become functional equations for the probability distribution of the effective field. These are not easy to solve, and available analytical results in this direction are mostly restricted to one dimension, or to models with infinite-ranged interactions. On the Bethe lattice, for short-ranged interactions with quenched disorder, e.g. in the prototypical case of the $\pm J$ random-exchange Ising model, the average free energy is trivially determined in the high temperature phase, but not in the low-temperature phase. It has not been possible so far to determine even the ground-state energy exactly despite several attempts.

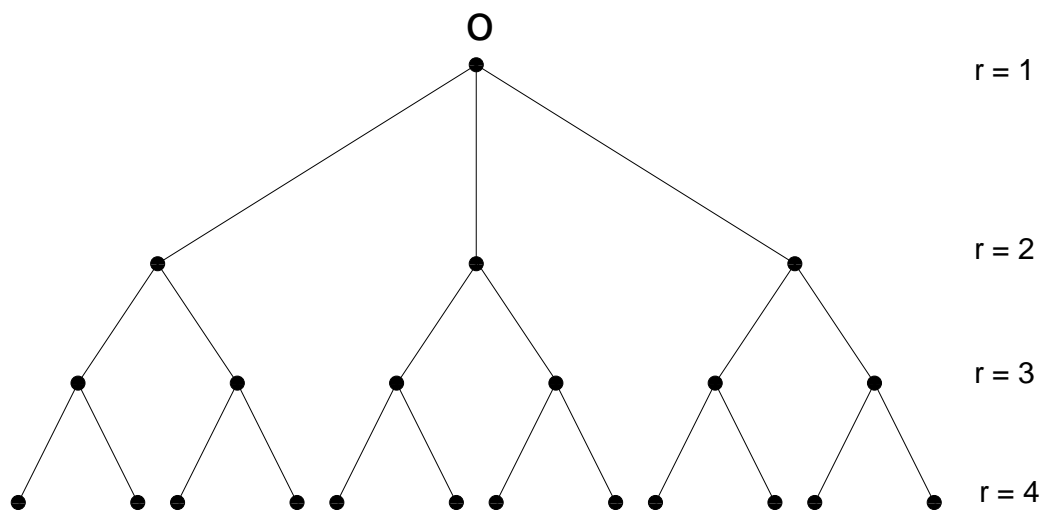

**Figure 2.4:** A Cayley tree of coordination number 3 and 4 generations.

The random field Ising model model on a Bethe lattice is special in that the zero-temperature *nonequilibrium* response to a slowly varying magnetic field can be determined exactly (Dhar et al. 1997). To be precise, the average non-equilibrium magnetization in this model can be determined if the magnetic field is increased very slowly, from $-\infty$ to $+\infty$, in the limit of zero temperature. It thus provides a good pedagogical model to study the slow relaxation to equilibrium in glassy systems.

The usual way to solve a problem on Bethe lattice is to consider the problem on a Cayley tree and calculate all the thermodynamic quantities in the deep inside of the tree. Consider a uniform Cayley tree of $n$ generations where each non-boundary site has a coordination number $z$ and the boundary sites have coordination number 1 [see Fig. 2.4]. The first generation consists of a single vertex. The $r$-th generation has $z(z-1)^{r-2}$ vertices for $r \geq 2$. At each vertex there is a Ising spin.

Because of the return point memory, to find the magnetization at field $h$ in the lower half of the hysteresis loop, we start with $h = -\infty$, when all spins are down and increase the field to $h$ in a single step. Now at this field, since the spins can be relaxed in any order (abelian property), we relax them in this: First all the spins at generation $n$ (the leaf nodes) are relaxed. Then spins at generation $n - 1$ are examined, and if any has a positive local field, it is flipped. Then we examine the spins at generation $n - 2$, and so on. If any spin is flipped, its descendent are reexamined for possible flips[†]. In this process, clearly the flippings of different spins of the same generation $r$ are independent events.

Let $P^{(r)}(h)$ be the probability that a spin on the $r$-th generation will be flipped when its parent spin at generation $r - 1$ is kept down, the external field is $h$, and each of its descendent spins has been relaxed. As each of the $z - 1$ direct descendents of a spin is independently up with probability $P^{(r+1)}$, the probability that exactly $m$ of them are up is $\binom{z-1}{m}[P^{(r+1)}]^m[1 - P^{(r+1)}]^{z-1-m}$. Suppose we pick a site at random in the tree away from the boundary, the probability that the local field at this site is positive, given that exactly $m$ of its neighbors are up, is precisely the probability that the local field $h_i$ at this site exceeds $[(z - 2m)J - h]$. We denote this probability by $p_m(h)$. Clearly,

$$p_m(h) = \int_{(z-2m)J-h}^{\infty} \phi(h_i)\, dh_i. \tag{2.7}$$

Now it is straightforward to write down a recursion relation for $P^{(r)}$ in terms of $P^{(r+1)}$:

$$P^{(r)}(h) = \sum_{m=0}^{z-1} \binom{z-1}{m} \left[P^{(r+1)}(h)\right]^m \left[1 - P^{(r+1)}(h)\right]^{z-1-m} p_m(h). \tag{2.8}$$

Given a value of $h$, we can determine $p_m(h)$ using Eq. (2.7). Then using Eq. (2.8), and the initial condition $P^{(n)} = p_0(h)$, $P^{(r)}$ can be determined for all $r < n$. For $r \ll n$, these

---

[†]This step is not really necessary if we are only interested in determining the magnetization at the site $O$. Skipping this step leads to considerable simplification of the relaxation process: First the spins of generation $n$ are examined, then those of $(n - 1)$ etc. till we finally examine the spin at $O$. No spin is checked more than once. The resulting configuration is not fully relaxed, but it is easy to prove that further relaxation will not change the state of the spin at $O$. The argument can be extended to show that the probability that an avalanches starting at $O$ is of size $s$ also is the same in this partially relaxed state as in the fully relaxed state.

probabilities tend to limiting value, $\lim_{n\to\infty} P^{(r)} = P^\star$, which satisfies the equation

$$P^\star(h) = \sum_{m=0}^{z-1} \binom{z-1}{m} [P^\star(h)]^m [1-P^\star(h)]^{z-1-m} \, p_m(h). \qquad (2.9)$$

This is a polynomial equation in $P^\star(h)$, which can be solved in terms of $\{p_m(h)\}$. Finally, for the spin at $O$, there are $z$ downward neighbors, and the probability that it is up is given by

$$\text{Prob}(s_O = +1 | h) = \sum_{m=0}^{z} \binom{z}{m} [P^\star(h)]^m [1-P^\star(h)]^{z-m} \, p_m(h). \qquad (2.10)$$

Because all spins deep inside the tree are equivalent, $\text{Prob}(s_O = +1 | h)$ determines the average magnetization for all sites deep inside the tree. This determines the lower half of the hysteresis loop. The upper half is obtained similarly.

For the three coordinated ($z=3$) Bethe lattice, the self-consistent equation satisfied by $P^\star$ [Eq. (2.9)] is quadratic and from physical arguments at least one root must vary between 0 and 1 continuously with $h$ for any value of disorder strength $\Delta$. Hence the magnetization given by Eq. (2.10) is also a continuous function of $h$. This is also the case with a linear chain ($z=2$), where the self-consistent equation [Eq. (2.9)] is linear.

On the other hand, the situation is quite different for $z \geq 4$. For example, for $z=4$, Eq. (2.9) is cubic, which has either one or three real roots which will vary with $h$. Figure 2.5 shows this variation for two values of disorder of the random field distribution given by

$$\phi(h_i) = \frac{1}{2\Delta} \text{sech}^2(h_i/\Delta). \qquad (2.11)$$

Note that for large disorder, there is only one real root which vary continuously from 0 to 1, giving rise to a continuous magnetization curve as shown in Fig. 2.6. But for small disorder, $P^\star(h)$ as a function of $h$ shows a "S" shaped curve, where at some value of $h$, two real root merge to becomes imaginary and disappear from the real plane. Therefore, as we vary $h$, on the physical ground initially $P^\star(h)$ takes the lower value till the point where it becomes complex and at that point it jumps to the upper value, giving rise to a jump discontinuity in the corresponding magnetization curve [Fig. 2.6].

This can be generalized to higher coordination number, where the mechanism of two real solutions of the polynomial equation Eq. (2.9) merging and both becoming complex is still the same.

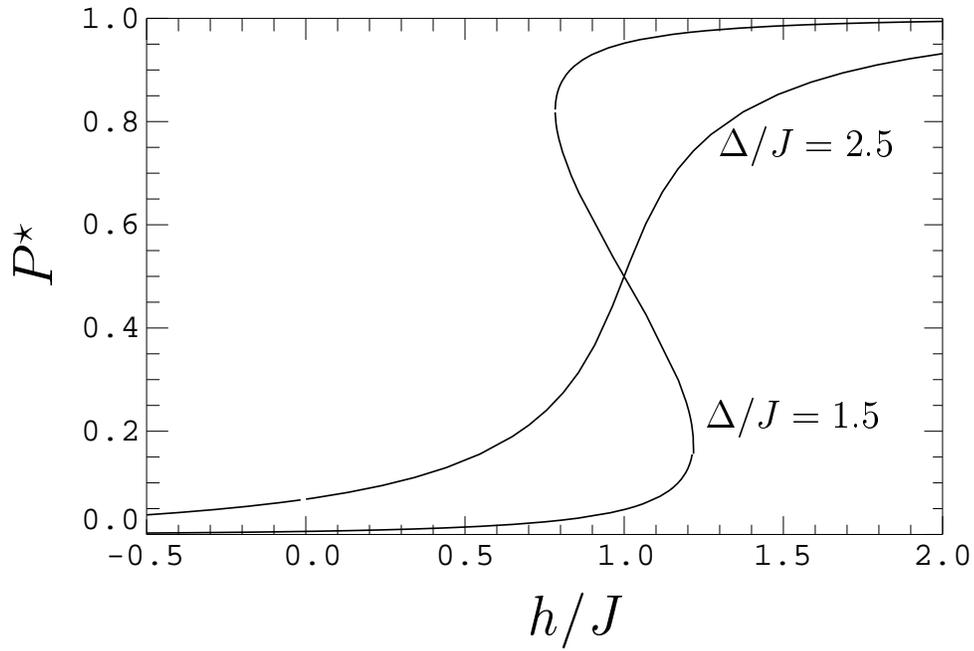

**Figure 2.5:** Variation of $P^\star(h)$ with $h$ for the Bethe lattice with $z = 4$, and the random field distribution given by Eq. (2.11).

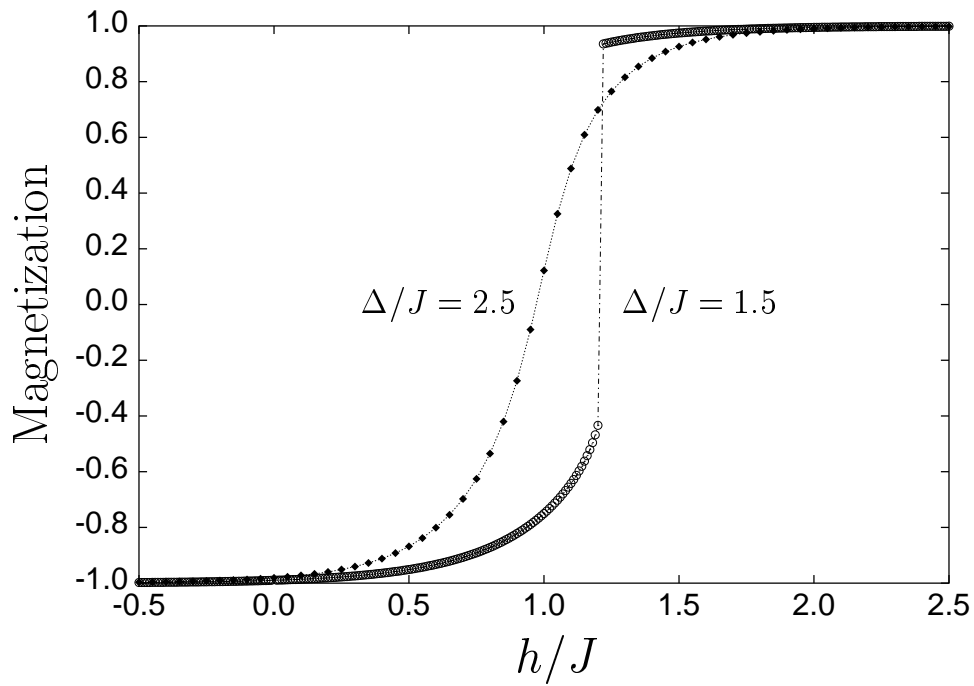

**Figure 2.6:** Magnetization as a function of increasing field for the Bethe lattice with $z = 4$ and the random field distribution given by Eq. (2.11).

# Chapter 3

# Distribution of avalanche sizes on the Bethe lattice

In this chapter, we study the distribution of avalanche sizes in the random field Ising model on a Bethe lattice (Sabhapandit et al. 2000). This chapter is organized as follows. In section 3.1, we set up a self-consistent equation for the generating function $Q(x)$ of the probability $Q_n$, that an avalanche propagating in subtree flips exactly $n$ more spins in the subtree before stopping, for arbitrary distribution of quenched random fields. Then we expressed the generating function $G(x)$ of distribution of avalanche sizes, in terms of $Q(x)$. In section 3.2, we consider the special case of a rectangular distribution of the random field. In this case, we explicitly solve the self-consistent equations for Bethe lattices with coordination numbers $z = 2$ and $3$. However, this case is non-generic. For small strength of disorder, the magnetization jumps from $-1$ to $+1$ at some value of the field, but for larger disorder, when the system shows finite avalanches, there is no jump in magnetization and the distribution function decays exponentially for large $s$. In section 3.3, we analyse the self-consistent equations to determine the form of the avalanche distribution for unimodal continuous distributions of the random field. We find that for coordination number $z \geq 4$, the magnetization shows a first order jump discontinuity as a function of the applied field at some field-strength $h_{\text{disc}}$, for weak disorder. Just below $h = h_{\text{disc}}$, the avalanche distribution has a universal $(-3/2)$ power-law tail.

## 3.1  Generating function for avalanche distribution

Consider a Cayley tree rooted at $O$, of $N$ generation [Fig. 2.4]. We will be interested in the portion of the tree where generation $r \ll N$, in the limit $N \to \infty$. Now consider the state of the system at external field $h$, and all the flippable sites have been flipped. We increase the field by a small amount $dh$ till one more site becomes unstable. We would like to calculate the probability that this would cause an 'avalanche' of $n$ spin flips. Since all sites deep inside are equivalent, we may assume the new susceptible site is the site $O$.

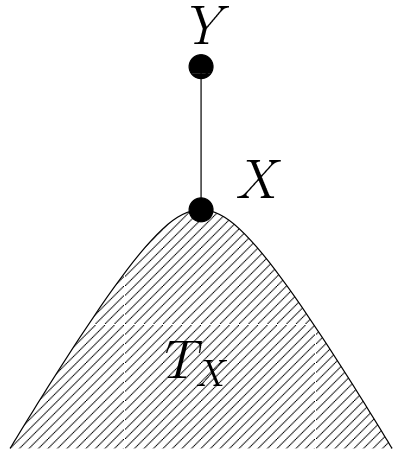

**Figure 3.1:** A sub-tree $T_X$ formed by $X$ and its descendents. The sub-tree is rooted at $X$ and $Y$ is the parent spin of $X$.

It is easy to see that this avalanche propagation is somewhat like propagation of infection in the contact process on the Bethe lattice. The 'infection' travels *downwards* from the site $O$ which acts as the initiator of infection. If any site is infected, then it can cause infection of some of its descendents. If the descendent spin is already up, it cannot be flipped; such sites act as immune sites for the infection process. If the descendent spin is down, it can catch infection with a finite probability. Furthermore, this probability does not depend on whether the other 'sibling' sites catch infection. Infection of two or more descendents of an infected site are uncorrelated events. Thus, we can expect to find the distribution of avalanches on the Bethe lattice, as for the size distribution of percolation clusters on a Bethe lattice (Stauffer and Aharony 1992). However, a precise description in terms of the contact process is complicated, as here the infection spreads in a correlated background of 'immune' (already up) spins, and the probability that a site catches infection does depend on the number of its neighbors that are already up.

We start with the initial configuration of all spins down. Now increase the external field to the value $h$. Consider a site $X$ at some generation $r > 1$ of the Cayley tree [Fig. 3.1]. We call the subtree formed by $X$ and its descendents $T_X$, the subtree rooted at $X$. We keep its parent spin $Y$ at generation $r - 1$ down, and relax all the sites in $T_X$ at the uniform field $h$. If $X$ is far away from the boundary, the probability that spin at $X$ is up is $P^\star(h)$, which is obtained by solving the self-consistent equation given by Eq. (2.9). The conditional probability that spin at a descendant of $X$ is up, given that the spin at $X$ is down is also $P^\star(h)$. We measure the response of $T_X$ to external perturbation by forcibly flipping the spin at $Y$ (whatever the local field there) and see how many spins in this subtree flip in response to this perturbation. Let $Q_n$ be the probability that the spin at $X$ was down when $Y$ was down *and* $n$ spins on the subtree $T_X$ flip up if $S_Y$ is flipped up. Here allowed values of $n$ are $0, 1, 2, \ldots$. Clearly, we have

$$P^\star + \sum_{n=0}^{\infty} Q_n = 1. \tag{3.1}$$

We define $Q(x)$ be the generating function of $Q_n$ as,

$$Q(x) = \sum_{n=0}^{\infty} Q_n x^n. \tag{3.2}$$

Clearly,

$$Q(x=0) = Q_0, \tag{3.3}$$
$$Q(x=1) = 1 - P^\star. \tag{3.4}$$

It is straight forward to write the self-consistent equation for $Q(x)$. Let us first relax all spins on $T_X$ keeping $X$ and $Y$ down. The probability that exactly $m$ the descendents of $X$ are turned up in this process be denoted by $\Pr(m)$. Clearly

$$\Pr(m) = \binom{z-1}{m} P^{\star m} (1 - P^\star)^{z-1-m}. \tag{3.5}$$

For a given $m$, the conditional probability that local field at $X$ is such that spin remains down, even if $Y$ is turned up is $1 - p_{m+1}$. Summing over $m$, and using the expression for $\Pr(m)$ above, we get

$$Q_0 = \sum_{m=0}^{z-1} \binom{z-1}{m} P^{\star m} (1 - P^\star)^{z-1-m} [1 - p_{m+1}]. \tag{3.6}$$

We can write down an expression for $Q_1$ similarly. In this case, if $m$ of the direct descendents of $X$ are up when $Y$ is down, the local field at all the remaining $z-1-m$ direct descendents must be such that they remain down even if $X$ is flipped up. This probability is $\binom{z-1}{m} P^{\star m} Q_0^{z-1-m}$. The local quenched field at $X$ must satisfy $(z-2m)J - h > h_X > (z-2m-2)J - h$. The probability for this to occur is $p_{m+1} - p_m$. Hence we get

$$Q_1 = \sum_{m=0}^{z-1} \binom{z-1}{m} P^{\star m} Q_0^{z-1-m} (p_{m+1} - p_m). \tag{3.7}$$

The equation determines $Q_n$ for higher $n$ can be written down similarly. It only involves the probabilities $Q_m$ with $m < n$ for the descendent spins. Formally we can write

$$Q_n = \sum_{m=0}^{z-1} \binom{z-1}{m} P^{\star m} \left[ \sum_{\{n_i\}=0}^{\infty} \prod_{i=1}^{z-1-m} Q_{n_i} \delta(\sum n_i, n-1) \right] (p_{m+1} - p_m), \quad (3.8)$$

where

$$\delta(i,j) \equiv \begin{cases} 1 & \text{for } i = j \\ 0 & \text{for } i \neq j \end{cases}$$

is the Kroneker delta.

These recursion equations are expressed more simply in terms of the generating function $Q(x)$. Multiplying both sides by $x^n$ and then summing over $n$, we see that the self-consistent equation for $Q(x)$ is

$$Q(x) = Q(x=0) + x \sum_{m=0}^{z-1} \binom{z-1}{m} P^{\star m} Q(x)^{z-1-m} (p_{m+1} - p_m). \quad (3.9)$$

This is a polynomial equation in $Q(x)$ of degree $z-1$, whose coefficients are functions of $h$ through $P^\star(h)$ and $p_m(h)$. Using Eq. (2.9) and Eq. (3.6), it is easily checked that for $x = 1$, the ansatz $Q(x=1) = 1 - P^\star$ satisfies the equation, as it should. To determine $Q(x)$ for any given external field $h$, we have to first solve the self-consistent equation for $P^\star$ [Eq. (2.9)]. This then determines $Q(x=0)$ using Eq. (3.6), and then, given $P^\star$ and $Q(0)$, we solve for $Q(x)$ by solving the $(z-1)$-th degree polynomial equation Eq. (3.9).

Finally, we express the relative frequency of avalanches of various sizes when the external field is increased from $h$ to $h + dh$ in terms of $Q(x)$. Let $G_s(h) \, dh$ be the probability that avalanche of size $s$ is initiated at $O$. We also define the generating function $G(x|h)$ as

$$G(x|h) = \sum_{s=1}^{\infty} G_s(h) x^s. \quad (3.10)$$

Consider first the calculation of $G_s(h)$ for $s = 1$. Let the number of descendents of $O$ that are up at field $h$ be $m$. For the spin at site $O$ to be down at $h$, but flip up at $h + dh$, the local field $h_O$ must satisfy $[(z - 2m)J - (h + dh)] < h_O < [(z - 2m)J - h]$. This occurs with probability $\phi(zJ - 2mJ - h) \, dh$. Each of the $(z - m)$ down neighbors of $O$ must not flip up, even when $s_O$ flips up. The conditional probability of this event is $Q_0^{z-m}$. Multiplying by the probability that $m$ neighbors are up, we finally get

$$G_1(h) = \sum_{m=0}^{z} \binom{z}{m} P^{\star m} Q_0^{z-m} \phi(zJ - 2mJ - h). \quad (3.11)$$

Arguing similarly, we can write the equation for $G_s(h)$ for $s = 2, 3$ etc. These equations simplify considerably when expressed in terms of the generating function $G(x|h)$, and we get

$$G(x|h) = x \sum_{m=0}^{z} \binom{z}{m} P^{\star m} Q(x)^{z-m} \phi(zJ - 2mJ - h). \tag{3.12}$$

In numerical simulations, and experiments, it is much easier to measure the avalanche distribution integrated over the full hysteresis loop. To get the probability that an avalanche of size $s$ will be initiated at any given site $O$ in the interval when the external field is increased from $h_1$ to $h_2$, we just have to integrate $G(x|h)$ in this range. For any $h$, the value of $dG/dx$ at $x = 1$ is proportional to the mean size of an avalanche, and thus to the average slope of the hysteresis loop at that $h$.

## 3.2  Explicit calculation for the rectangular distribution

While the general formalism described in the previous section can be used for any distribution, and any coordination number, to calculate the avalanche distributions explicitly, we have to choose some specific form for the probability distribution function. In this section, we shall consider the specific choice of a rectangular distribution: The quenched random field is uniformly distributed between $-\Delta$ and $\Delta$, so that

$$\phi(h_i) = \frac{1}{2\Delta}, \quad \text{for} \quad -\Delta \leq h_i \leq \Delta. \tag{3.13}$$

In this case, the cumulative probabilities $p_m(h)$ become piece wise linear functions of $h$, and $h$-dependence of the distribution is easier to work out explicitly. We shall work out the distributions for the linear chain ($z = 2$), and the 3-coordinated Bethe lattice.

### 3.2.1  The linear chain ($z = 2$)

The simplest illustration is for a linear chain. In this case the self-consistent equation, for the probability $P^\star$ [Eq. (2.9)] becomes a linear equation. This is easily solved, and explicit expressions for $Q_0$, and $Q(x)$ are obtained (see Appendix A.1). The different regimes showing different qualitative behavior of the hysteresis loops are shown in Fig. 3.2

For $h < 2J - \Delta$ (region A), all the spin remain down. For $h > \Delta$, all spins are up (region D). For $\Delta < J$, we get a rectangular loop and the magnetization jumps discontinu-ously from $-1$ to $+1$ in a single infinite avalanche, and we directly go from region A to D as the field is increased. For $\Delta > J$, we get nontrivial hysteresis loops.

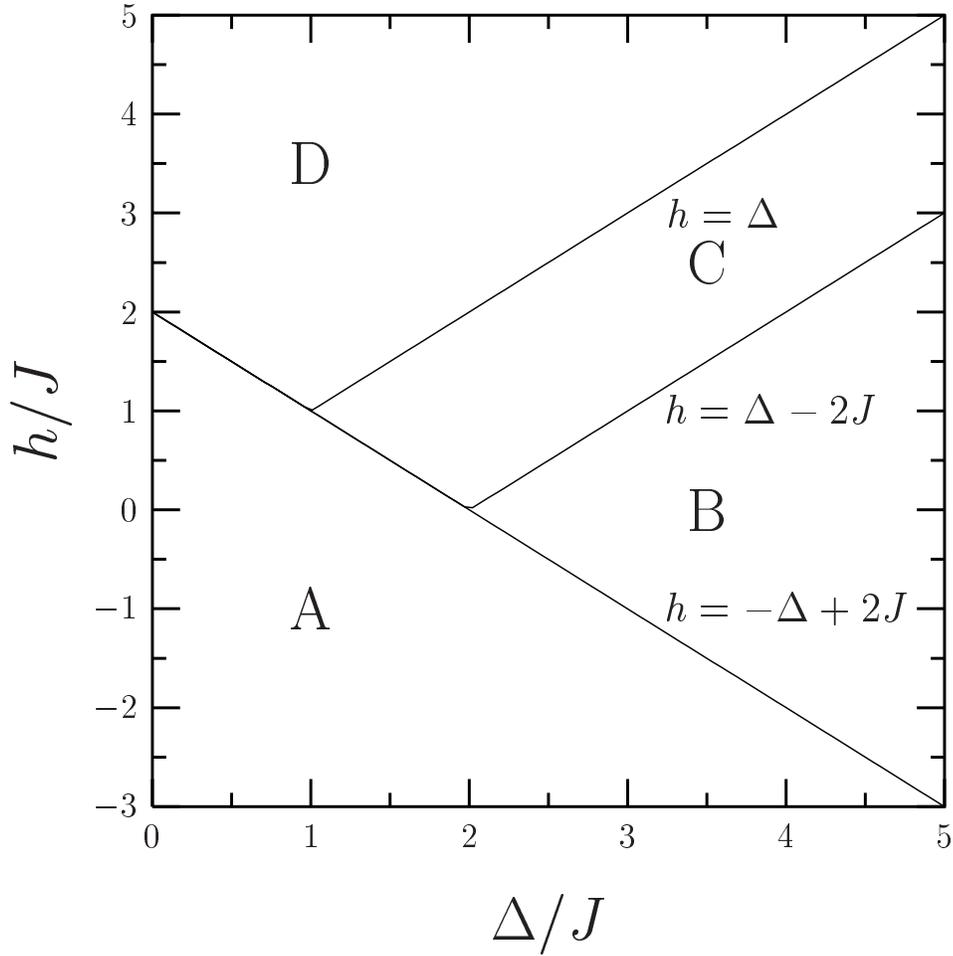

**Figure 3.2:** Behavior of RFIM in the magnetic field - disorder ($h - \Delta$) plane for a linear chain. The regions A-D correspond to qualitatively different responses. In region A all spins are down and in region D all are up. The avalanches of finite size occur in region B and C.

The hysteresis loops for different values of $\Delta = 0.5, 1.5$ and $2.5$ are shown in Fig. 3.3. If $\Delta$ is sufficiently large ($\Delta > J$), we find that the mean magnetization is a precisely linear function of the external field for a range of values of the external field $h$ (region B in Fig. 3.2). For larger $h$ values, the magnetization shows saturation effects, and is no longer linear (region C).

The explicit forms of the generating function $Q(x)$ are given in the Appendix A.1. We find that in region B, the function $Q(x)$ is independent of the applied field $h$. The

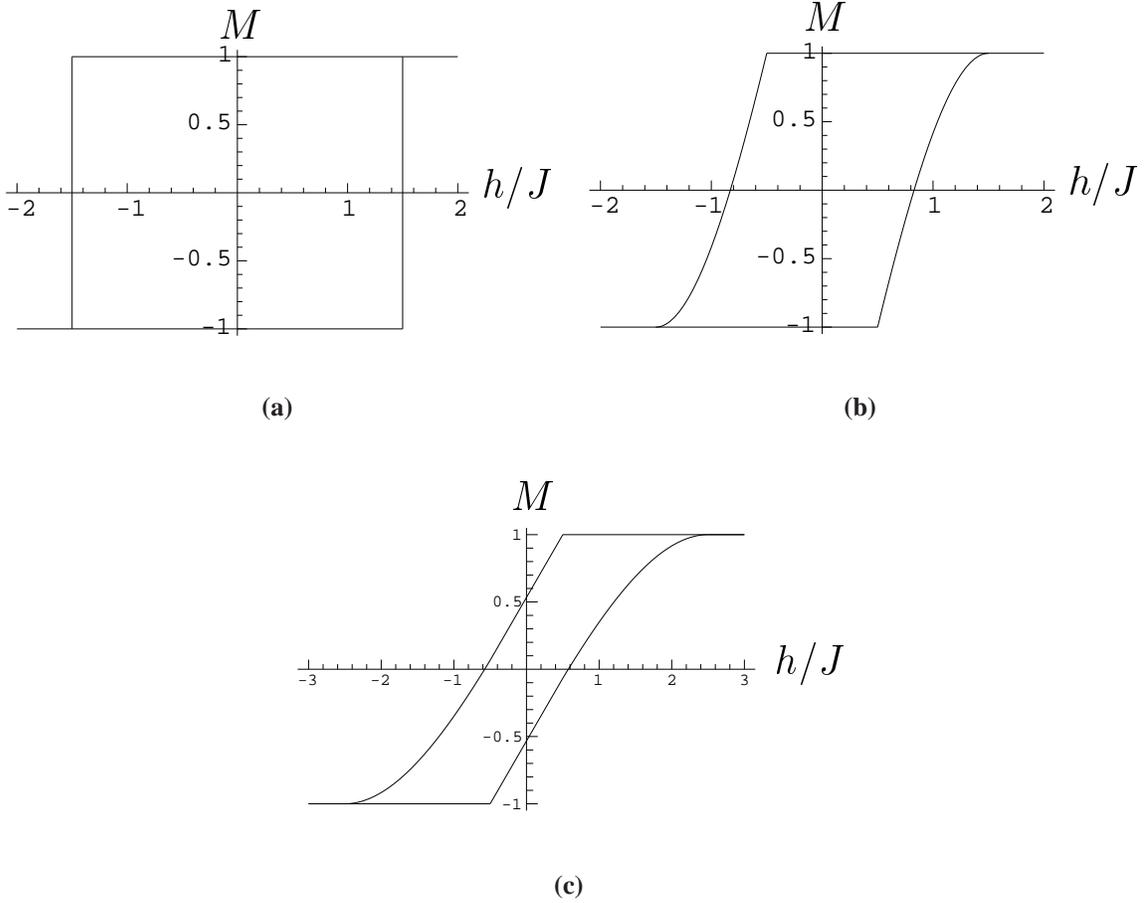

**Figure 3.3:** Hysteresis loops for the linear chain for the rectangular distribution of quenched fields with different widths. (a) $\Delta/J = 0.5$, (b) $\Delta/J = 1.5$ and (c) $\Delta/J = 2.5$

distribution function $G_s(h)$ has a simple dependence on $s$ of the form

$$G_s(h) = A_1 s \left(\frac{J}{\Delta}\right)^s, \tag{3.14}$$

where $A_1$ is a constant, that depends only on $J/\Delta$, and does not depend on $s$ or $h$,

$$A_1 = \frac{1}{2\Delta} \frac{(1 - J/\Delta)^2}{(J/\Delta)}. \tag{3.15}$$

In region C, the mean magnetization is a nonlinear function of $h$. But $Q(x)$ is still a rational function of $x$. From the explicit functional form of $Q(x)$ and $G(x|h)$ are given in the appendix A.1, we find that $G_s(h)$ is of the form

$$G_s(h) = [A'_1 s + A'_2] \left(\frac{J}{\Delta}\right)^s, \quad \text{for} \quad s \geq 2. \tag{3.16}$$

Here $A_1'$ and $A_2'$ have no dependence on $s$ but are explicit functions of $h$.

Integrating over $h$ from $-\infty$ to $\infty$ we get the integrated avalanche distribution $D_s$,

$$D_s = \int_{-\infty}^{\infty} G_s(h)\, dh. \qquad (3.17)$$

It is easy to see from above that the integrated distribution $D_s$ also has the form

$$D_s = [A_2 s + B_2]\left(\frac{J}{\Delta}\right)^s, \quad \text{for} \quad s \geq 2, \qquad (3.18)$$

where the explicit forms of the coefficients $A_2$ and $B_2$ are given in the Appendix A.1.

### 3.2.2 The case $z = 3$

The analysis for the case $z = 3$ is very similar to the linear case. In this case, the self-consistent equation for $P^\star(h)$ [Eq. (2.9)] becomes a quadratic equation. The qualitative behavior of solution is very similar to the earlier case. Some details are given in Appendix A.2. We again get regions A-D as before, but the boundaries are shifted a bit, and are shown in Fig. 3.4. As before, in region B, the average magnetization is a linear function of $h$, and the avalanche distribution is independent of $h$.

We find that in regime B, the distribution of avalanche sizes is given by

$$G_s(h) = N \left[\frac{(2s)!}{(s-1)!(s+2)!}\right](1 - J/\Delta)^s \left(\frac{J}{\Delta}\right)^s, \qquad (3.19)$$

where $N$ is a normalization constant given by

$$N = \frac{3}{2\Delta}(1 - J/\Delta)^2 \frac{1}{(J/\Delta)}. \qquad (3.20)$$

It is easy to see that for large $s$, $G_s(h)$ varies as

$$G_s \sim s^{-\frac{3}{2}} \kappa^s, \qquad (3.21)$$

where

$$\kappa = 4(1 - J/\Delta)(J/\Delta). \qquad (3.22)$$

In region B, $J/\Delta$ is always less than $1/3$, and so this function always has an exponential decay for large $s$.

In the region C, we find that the avalanche distribution is of the form

$$G_s(h) = N' \left[\frac{(2s)!}{(s-1)!(s+2)!}\right]\kappa^s, \qquad (3.23)$$

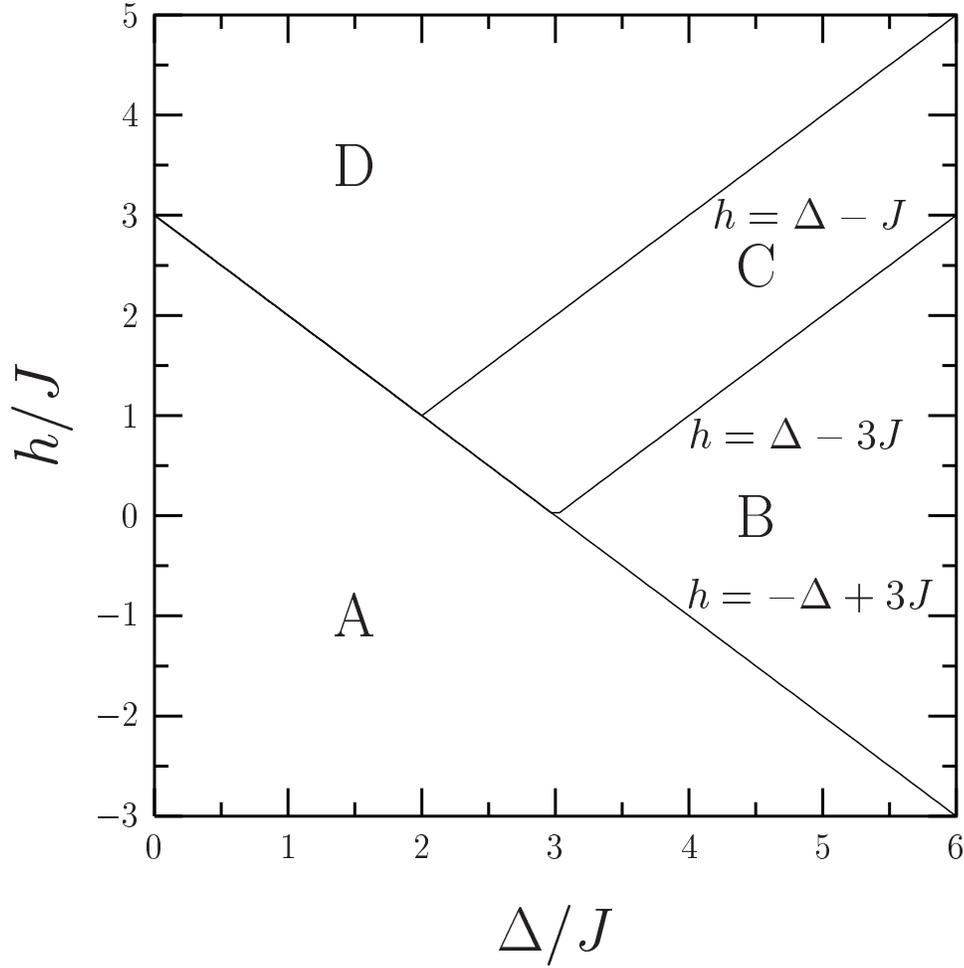

**Figure 3.4:** Behavior of RFIM in the magnetic field - disorder ($h - \Delta$) plane for Bethe lattice of coordination number 3. The qualitative behavior in different regions A-D is similar to that of a linear chain (Fig. 3.2).

where $N'$ is a normalization constant independent of $s$, and $\kappa$ is a cubic polynomial in the external field $h$:

$$\kappa = \frac{1}{8(1 - 2J/\Delta)^2} \left[ \left\{ 9 - 53(J/\Delta) + 119(J/\Delta)^2 - 107(J/\Delta)^3 \right\} \right.$$
$$+ \left\{ -5 + 10(J/\Delta) + 11(J/\Delta)^2 \right\} (h/\Delta)$$
$$\left. + \left\{ 3 - 9(J/\Delta)^2 \right\} (h/\Delta)^2 + (h/\Delta)^3 \right]. \tag{3.24}$$

For any fixed $s$, the integrated distribution $D_s$ can be evaluated explicitly, but become lengthy even for small $s$.

## 3.3 General distributions

The analysis of the previous section can, in principle, be extended to higher coordination numbers, and other distributions of random fields. However, the self-consistent equations become cubic, or higher order polynomials. In principle, an explicit solution is possible for $z \leq 5$, but it is not very instructive. However, the qualitative behavior of solutions is easy to determine, and is the same for all $z \geq 4$. We shall take $z = 4$ in the following for simplicity. Since we only study the general features of the self-consistent equations, we need not pick a specific form for the continuous distributions of random field distribution $\phi(h_i)$. We shall only assume that it has a single maximum around zero and asymptotically go to zero at $\pm\infty$, as shown in Fig. 3.5.

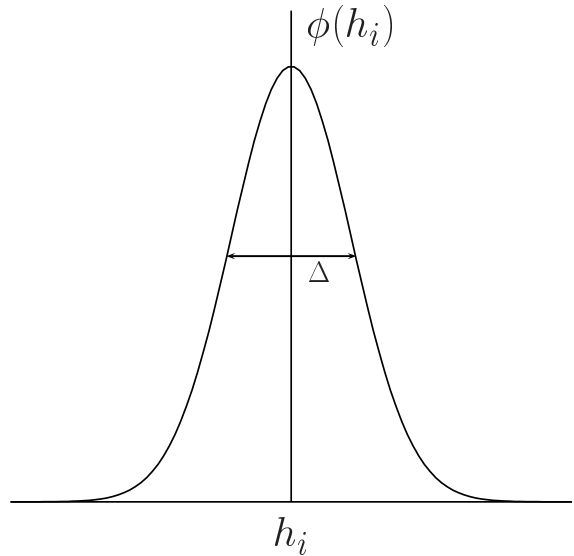

**Figure 3.5:** A schematic plot of a unimodal random field distribution which asymptotically go to zero at $\pm\infty$.

For small width ($\Delta$) of the random field distribution i.e. for weak disorder the magnetization shows a jump discontinuity as a function of the external uniform field, which disappears for a larger values of $\Delta$ (section 2.4). For fields $h$ just lower than the value where the jump discontinuity occurs, the slope of the hysteresis curves is large, and tends to infinity as the field tends to the value at which the jump occurs. This indicates that large avalanches are more likely just before the first order jump in magnetization.

For $z = 4$, the self-consistent equation for $P^\star(h)$ [Eq. (2.9)] is cubic

$$aP^{\star 3} + bP^{\star 2} + cP^\star + d = 0 \tag{3.25}$$

where $a, b, c$ and $d$ are functions of the external field $h$, expressible in terms of the cumulative probabilities $p_i, i = 0$ to $3$,

$$a = p_3 - 3p_2 + 3p_1 - p_0, \qquad (3.26a)$$

$$b = 3p_2 - 6p_1 + 3p_0, \qquad (3.26b)$$

$$c = 3p_1 - 3p_0 - 1, \qquad (3.26c)$$

$$d = p_0. \qquad (3.26d)$$

This equation will have 1 or 3 real roots, which will vary with $h$. We have shown this variation for the real roots which lie between 0 and 1 in Fig. 2.5 for the case where $\phi(h_i)$ is a simple distribution given by Eq. (2.11).

We have also solved numerically the self-consistent equation for $P^\star$ for other choices of $p(h_i)$, like the gaussian distribution, and for higher $z (= 4, 5, 6)$. In each case we find that the qualitative behavior of the solution is very similar. Note that the rectangular distribution discussed in the previous section is very atypical in that both the coefficients $a$ and $b$ vanish for an entire range of values of $h$.

In the generic case, we find two qualitatively different behaviors: For larger values of $\Delta$, there is only one real root for any $h$. For $\Delta$ sufficiently small, we find a range of $h$ where there are 3 real solutions. There is a critical value $\Delta_c$ of the width which separates these two behaviors. For the particular distribution chosen [Eq. (2.11)], $\Delta_c \simeq 2.10382$.

In the first case, the real root is a continuous function of $h$, and correspondingly, the magnetization is a continuous function of $h$. This is the case corresponding to $\Delta = 2.5$ in Fig. 2.5.

For smaller $\Delta < \Delta_c$, for large $\pm h$ there is only one root, but in the intermediate region there are three roots. The typical variation is shown for $\Delta = 1.5$ in Fig. 2.5. In the increasing field the probability $P^\star(h)$ initially takes the smallest root. As $h$ increases, at a value $h = h_{\text{disc}}$, the middle and the lower roots become equal and after that both disappear from the real plane. At $h = h_{\text{disc}}$ the probability $P^\star(h)$ jumps to the upper root. Thus for $\Delta < \Delta_c$ there is a discontinuity in $P^\star(h)$ which gives rise to a first order jump in the magnetization curve.

The field $h_{\text{disc}}$ where the discontinuity of magnetization occurs, is determined by the condition that for this value of $h$, the cubic equation [Eq. (3.25)] has two equal roots. The value of $P^\star$ at this point, denoted by $P^\star_{disc}$, satisfies the equation

$$3a_0 P^{\star 2}_{disc} + 2b_0 P^\star_{disc} + c_0 = 0, \qquad (3.27)$$

where $a_0, b_0$ and $c_0$ are the values of $a, b$ and $c$ at $h = h_{\text{disc}}$.

We now determine the behavior of the avalanche generating function $G_s(h)$ for large $s$ and $h$ near $h_{\text{disc}}$. The behavior for large $s$ corresponds to $x$ near 1. So we write $x = 1 - \delta$, with $\delta$ small, and $h = h_{\text{disc}} - \epsilon$. Near $h_{\text{disc}}$, $a, b, \ldots$ vary linearly with $\epsilon$ and

$$P^\star \approx P^\star_{disc} - \alpha\sqrt{\epsilon} + O(\epsilon), \tag{3.28}$$

where $\alpha$ is a numerical constant.

Since $Q(x = 1) = 1 - P^\star(h)$, if $x$ differs slightly from unity $Q(x)$ also differs from $1 - P^\star(h)$ by a small amount. Substituting $x = 1 - \delta$ and $Q(x = 1 - \delta) = 1 - P^\star - F(\epsilon, \delta)$ in the self-consistent equation for $Q(x)$ [Eq. (3.9)], where both $\delta$ and $F$ are small, using Eq. (3.27), we get to lowest order in $\delta$, $\epsilon$ and $F$

$$F^2 + \beta\sqrt{\epsilon}F - \gamma^2\delta = 0, \tag{3.29}$$

where $\beta$ and $\gamma$ are some constants. Thus, to lowest orders in $\epsilon$ and $\delta$, $F$ is given by

$$F = (1/2)\left[\sqrt{\beta^2\epsilon + 4\gamma^2\delta} - \beta\sqrt{\epsilon}\right]. \tag{3.30}$$

Thus $Q(x)$ has leading square root singularity at $x = 1 + \frac{\beta^2\epsilon}{4\gamma^2}$. Consequently, $G(x|h)$ will also show a square root singularity $x = 1 + \frac{\beta^2\epsilon}{4\gamma^2}$. This implies that the Taylor expansion coefficients $G_s(h)$ vary as

$$G_s(h) \sim s^{-\frac{3}{2}}\left(1 + \frac{\beta^2\epsilon}{4\gamma^2}\right)^{-s}, \quad \text{for large } s. \tag{3.31}$$

At $\epsilon = 0$, we get

$$G_s(h_{\text{disc}}) \sim s^{-\frac{3}{2}}. \tag{3.32}$$

Thus at $h = h_{\text{disc}}$ the avalanche distribution has a power law tail.

To calculate the integrated distribution $D_s$, we have to integrate Eq. (3.31) over a range of $\epsilon$ values. For large $s$, only $\epsilon < \frac{\gamma^2}{\beta^2 s}$ contributes significantly to the integral, and thus we get

$$D_s \sim s^{-\frac{5}{2}}, \quad \text{for large } s. \tag{3.33}$$

Thus the integrated distribution shows a robust $(-5/2)$ power law for a range of disorder strength $\Delta$.

# Minor hysteresis loops on the Bethe lattice

In this chapter, we derive exact self-consistent equations to obtain magnetization on the minor loops as a function of external field for arbitrary distribution of quenched random fields on a Bethe lattice. The return hysteresis loops for the linear chain was obtained by Shukla (2000). In sec. 2.4, we have discussed how to obtain the magnetization on the lower hysteresis curve, i.e. if we start with $h = -\infty$, when all the spins are down, and then slowly increase the external field. Now suppose the system is on the lower hysteresis curve at some external field $h_1$. Decreasing the field from $h_1$ to some field $h_2$ and then again increasing to $h_1$, we obtain the first minor loop. Similarly starting from the first minor loop at some field $h_3$ and decreasing the field to $h_4$, and then increasing to $h_3$, we obtain the second minor loop and so on. Figure 4.1 shows two minor loops. In general, the $n$-th minor loop for $n > 1$ is obtained from the lower half of $(n-1)$-th minor loop by decreasing the field from $h_{2n-1}$ to $h_{2n}$ and then increasing to $h_{2n+1} < h_{2n-1}$. This involves $\{h_n\} \equiv h_n, h_{n-1}, ..., h_1$, the history of all the turning points from $h_1$ to $h_n$. In the next section we will obtain the exact expressions for the magnetizations on the minor loops for arbitrary distributions of random fields. Similar results were later obtained independently by Shukla (2001).

## 4.1 Magnetization on minor loops

In sec. 2.4, we have determined the average magnetization in the deep inside the Cayley tree, on the lower hysteresis loop for arbitrary distributions of random field distributions. The average magnetization is equivalent to the magnetization at the root $O$ of the Cayley tree [Fig. 2.4], in the limit the number of generations, $n \to \infty$. We obtained the magnetization at $O$ as,

$$\operatorname{Prob}(s_O = +1|\, h) = \sum_{m=0}^{z} \binom{z}{m} [P^\star(h)]^m [1 - P^\star(h)]^{z-m} \, p_m(h), \tag{2.10}$$

where $P^\star(h)$ is the limiting value ($r \ll n$, and the limit $n \to \infty$) of conditional probability $P^{(r)}$, that a spin on the $r$-th generation will be flipped when its parent spin at generation $r-1$ is kept down, the external field is $h$, and each of its descendent spins has been relaxed, and is obtained by solving the polynomial equation

$$P^\star(h) = \sum_{m=0}^{z-1} \binom{z-1}{m} [P^\star(h)]^m [1 - P^\star(h)]^{z-1-m} \, p_m(h), \tag{2.9}$$

and $p_m(h)$ is the probability that a spin flips up, given that exactly $m$ neighbors are up, which is obtained by integrating the random field distribution $\phi(h_i)$ as,

$$p_m(h) = \int_{(z-2m)J-h}^{\infty} \phi(h_i)\, dh_i. \tag{2.7}$$

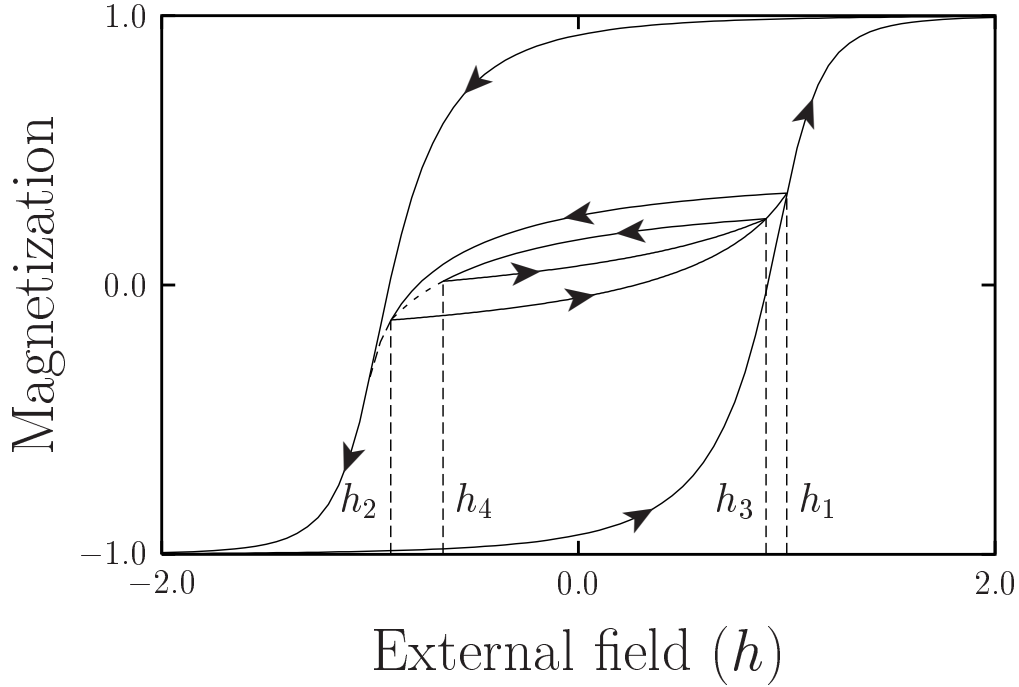

**Figure 4.1:** Minor hysteresis loops for Bethe lattice.

Similarly for the upper half of the hysteresis loop, when the external field is decreased from $\infty$, we can define $Q^{(r)}(h)$ to be the conditional probability that a spin on the $r$-th generation will be flipped down when its parent spin at generation $r - 1$ is kept up, the external field is decreased from $\infty$ to $h$, and each of its descendent spins has been relaxed. The limiting value $Q^\star(h)$ also satisfies self-consistent equation

$$Q^\star(h) = \sum_{m=0}^{z-1} \binom{z-1}{m} [1 - Q^\star(h)]^m [Q^\star(h)]^{z-1-m} [1 - p_{m+1}(h)], \qquad (4.1)$$

and in terms of $Q^\star(h)$ the upper half of the major loop can be obtained. Since $p_{m+1}(h) = p_m(h + 2J)$, the recursion relation satisfies by $1 - Q^{(r)}(h - 2J)$, is same as the relation satisfies by $P^{(r)}(h)$ which is given by Eq. (2.8). Therefore, we conclude that $Q^{(r)}(h-2J) = 1 - P^{(r)}(h)$.

### 4.1.1 First minor hysteresis loop

Suppose the system is on the lower hysteresis curve at some external field $h_1$. Now if the field is decreased from $h_1$ to some field $h_2$ and then again increased to $h_1$, return point memory [section 2.1] ensures that the loop closes. This is the first minor loop [see Fig. 4.1]. Now when the applied field is increased from $-\infty$ to $h_1$ and then decreased to a field $h_2$, to find out the spins which can flip down we need to consider only about the subset of spins which flipped up at field $h_1$. Suppose a spin at a randomly chosen site flips up at field $h_1$. As a result, the net local field at each of its nearest neighbors increases by an amount $2J$ and some of down neighbors might become unstable. We flip up those spins at time step 1. After flipping them more of their neighbor might become unstable. We flip them up in time step 2 and so on. This process will be continued till the avalanche stop. Figure 4.2 shows the order at which spins flip during a particular avalanche. Now in this avalanche if a spin $s_i$ flips up at time step $t$ and as a result, if $m$ of its neighbors flip at time step $t + 1$, then the local field at $i$ will increase by $2mJ$. Therefore when the field is decreased to $h_2 \geq h_1 - 2J$, $s_i$ can not flip back at $h_2$ unless all the neighbors which had flipped at time step $t + 1$ after $s_i$ flipped up, again flip back at $h_2$. Therefore, the spin which was the initiator of the avalanche (which flipped at time step 0) can flip down at $h_2$ only at the end, after all the spins of that avalanche flip back and in this flip-back avalanche the spins flip exactly in the reverse time order to the previous avalanche. This property will be called the *time ordering* property of spin-flip-back process.

Consider the case, when the system is on the lower half of the major loop at field $h_1$ and then the field is decreased to $h_1 - 2J$. Then all the neighbors of a vertex $i$ which had flipped up at $h_1$ after $s_i$ flipped up will flip back at $h_1 - 2J$, since they flipped up when their local fields had been increased by $2J$. Therefore, the conditional probability that a spin is down at $h_1 - 2J$, given its parent spin is up is same as the conditional probability that a spin is down at $h_1$, given its parent spin is down, which is $1 - P^{(r)}(h_1)$. The later is again equal to $Q^{(r)}(h_1 - 2J)$, the conditional probability that a spin is down when the field is decreased from $\infty$ to $h_1 - 2J$, given its parent spin is kept up. Therefore the reverse magnetization curve starting from $h_1$, meets the upper major half at $h_1 - 2J$ and merge with it for $h_2 < h_1 - 2J$. This result can be generalized for arbitrary graphs, which is discussed in section 4.2. Thus, we can consider the first minor loop in the range $[h_1 - 2J, h_1]$. Since in this range of external field the spin-flip-back process obeys *time ordering*, if a spin $s_i$ flips up at $h_1$ and flips down at $h_2$, then the probability that a neighbor of it at generation $r$ is up before $s_i$ flips back at $h_2$ is same as the neighbor was up before $s_i$ flipped up at $h_1$, given by $P^{(r)}(h_1)$. The probability that a neighbor is down before $s_i$ flips down at $s_i$ can be splitted into two parts:

(1) it didn't flip up after $s_i$ flipped up at $h_1$ and

(2) it flipped up after $s_i$ flipped up at $h_1$ and flips back at $h_2$ before $s_i$ flips back.

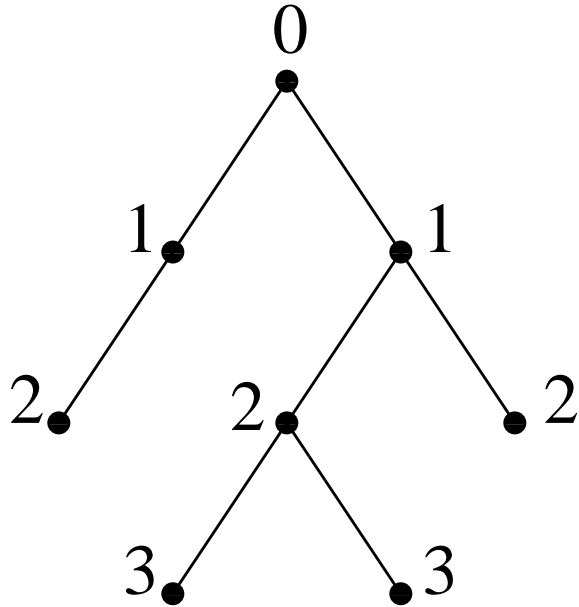

**Figure 4.2:** Time order at which spins flip during a particular avalanche.

Consider a site $X$ at some level $r$ of the Cayley tree [Fig. 3.1]. We call the subtree formed by $X$ and its descendents $T_X$, the subtree rooted at $X$. We keep its parent spin $Y$ at generation $r - 1$ down, and relax all the sites in $T_X$ at the uniform field $h_1$. Let $R_-^{(r)}(h_1)$ be the probability that $s_X$ remains down after $s_Y$ turned up at $h_1$. For $r \ll n$, in the limit

$n \to \infty$, these probabilities tends to limiting value $R_-^\star(h_1)$, given by

$$R_-^\star(h_1) = 1 - \sum_{m=0}^{z-1} \binom{z-1}{m} [P^\star(h_1)]^m [1-P^\star(h_1)]^{z-1-m} \, p_{m+1}(h_1). \qquad (4.2)$$

Let $G_-^{(r)}(h_2, h_1)$ be the conditional probability that:

(a) $s_X$ was down at $h_1$, given $s_Y$ was down,

(b) $s_X$ flipped up at $h_1$ after $s_Y$ flipped up and

(c) $s_X$ flips back at $h_2$, given $s_Y$ is still up.

Then a recursion relation for $G_-^{(r)}(h_2, h_1)$ in terms of $G_-^{(r+1)}(h_2, h_1)$ can be written as

$$G_-^{(r)}(h_2, h_1) = \sum_{m=0}^{z-1} \binom{z-1}{m} [P^\star(h_1)]^m \left[R_-^{(r+1)}(h_1) + G_-^{(r+1)}(h_2, h_1)\right]^{z-1-m}$$
$$\times [p_{m+1}(h_1) - p_{m+1}(h_2)], \qquad (4.3)$$

and its limiting value $G_-^\star(h_2, h_1)$, satisfies the self-consistent equation

$$G_-^\star(h_2, h_1) = \sum_{m=0}^{z-1} \binom{z-1}{m} [P^\star(h_1)]^m \left[R_-^\star(h_1) + G_-^\star(h_2, h_1)\right]^{z-1-m}$$
$$\times [p_{m+1}(h_1) - p_{m+1}(h_2)]. \qquad (4.4)$$

This is a polynomial equation in $G_-^\star(h_2, h_1)$ of degree $z-1$, whose coefficients are functions of $h_1$ and $h_2$ through $P^\star(h_1)$, $R_-^\star(h_1)$, $p_m(h_1)$ and $p_m(h_2)$. To determine $G_-^\star(h_2, h_1)$ for any given pair of external fields $h_1$ and $h_2$, we have to first solve the self-consistent equation for $P^\star(h_1)$ [Eq. (2.9)]. This then determines $R_-^\star(h_1)$ using Eq. (4.2), and then, given $P^\star(h_1)$ and $R_-^\star(h_1)$, we solve for $G_-^\star(h_2, h_1)$ by solving the $(z-1)$-th degree polynomial equation Eq. (4.4). Now the decrease in magnetization, when the field is decreased from $h_1$ to $h_2$, is determined by the probability that a spin at $O$ was up at $h_1$ and turns down at $h_2$, given by,

$$\text{Prob}(s_O = -1; h_2 \mid s_O = +1; h_1) =$$
$$\sum_{m=0}^{z} \binom{z}{m} [P^\star(h_1)]^m \left[R_-^\star(h_1) + G_-^\star(h_2, h_1)\right]^{z-m} [p_m(h_1) - p_m(h_2)]. \qquad (4.5)$$

This determines the upper half of first minor loop.

Similarly when the field is again reversed from $h_2$ to $h_3 < h_1$, using the symmetry between up an down spins it is easy to see that again the *time ordering* property holds. Therefore the probability that the neighbor of a spin $s_i$ is down before $s_i$ flips up at $h_3$ is $[R_-^{(r)}(h_1) + G_-^{(r)}(h_2, h_1)]$. The probability that a neighbor is up at $h_3$ before $s_i$ flips up is given by sum of two probabilities $R_+^{(r)}(h_2, h_1)$ and $G_+^{(r)}(h_3, h_2, h_1)$; where $R_+^{(r)}(h_2, h_1)$ is the probability that: (a) $s_X$ is up at $h_1$, given that $s_Y$ is kept down and $T_X$ is relaxed, (b) $s_Y$ flips up at $h_1$ and (c) $s_X$ remains up after $s_Y$ flips down at $h_2$ and $T_X$ is relaxed and $G_+^{(r)}(h_3, h_2, h_1)$ is the probability that: (a) $s_X$ is up at $h_1$, given that $s_Y$ is kept down and $T_X$ is relaxed, (b) $s_Y$ flipped up at $h_1$, (c) $s_X$ flipped down after $s_Y$ flipped down at $h_2$ and (d) $s_X$ flips back at $h_3$, given $s_Y$ is still down.

$R_+^{(r)}(h_2, h_1)$ is the equal to the probability that the spin is up at $h_1$ minus the probability that it becomes down at $h_2$. Its limiting value is given by

$$R_+^\star(h_2, h_1) = P^\star(h) - \sum_{m=0}^{z-1} \binom{z-1}{m} [P^\star(h_1)]^m \left[R_-^\star(h_1) + G_-^\star(h_2, h_1)\right]^{z-1-m}$$
$$\times [p_m(h_1) - p_m(h_2)]. \quad (4.6)$$

The limiting value $G_+^\star(h_3, h_2, h_1)$ satisfies the self-consistent equation

$$G_+^\star(h_3, h_2, h_1) = \sum_{m=0}^{z-1} \binom{z-1}{m} \left[R_+^\star(h_2, h_1) + G_+^\star(h_3, h_2, h_1)\right]^m$$
$$\times \left[R_-^\star(h_1) + G_-^\star(h_2, h_1)\right]^{z-1-m} [p_m(h_3) - p_m(h_2)]. \quad (4.7)$$

Solving the above self-consistent equation [Eq. (4.7)] we determine $G_+^\star(h_3, h_2, h_1)$ and then the increase in magnetization, when the field is increased from $h_2$ to $h_3$ is determined in terms of the following probability:

$$\text{Prob}(s_O = +1; h_3 \mid s_O = -1; h_2) =$$
$$\sum_{m=0}^{z} \binom{z}{m} \left[R_+^\star(h_2, h_1) + G_+^\star(h_3, h_2, h_1)\right]^m \left[R_-^\star(h_1) + G_-^\star(h_2, h_1)\right]^{z-m}$$
$$\times [p_m(h_3) - p_m(h_2)], \quad (4.8)$$

which determines the lower half of first minor loop.

### 4.1.2 General minor hysteresis loops

In the previous sub section, we obtained the first minor loop. The other minor loops can be obtained similarly. In all the minor loops the spin-flip-back process obeys *time ordering*. In general, the $n$-th minor loop for $n > 1$ is obtained from the lower half of $(n-1)$-th minor loop by decreasing the field from $h_{2n-1}$ to $h_{2n}$ and then increasing to $h_{2n+1} < h_{2n-1}$. For convenience, we will use the notation $\{h_n\} \equiv h_n, h_{n-1}, ..., h_1$ for the history of all the turning points from $h_1$ to $h_n$.

On the upper half of the $n$-th minor loop ($n > 1$), when the field is decreased from $h_{2n-1}$ to $h_{2n}$, the probability that a neighbor of a spin $s_i$ is up before $s_i$ (which is deep inside the tree) flips down at $h_{2n}$ is $[R^\star_+(\{h_{2n-2}\}) + G^\star_+(\{h_{2n-1}\})]$. The probability that a neighbor of $s_i$ is down before it flips down is given by $[R^\star_-(\{h_{2n-1}\}) + G^\star_-(\{h_{2n}\})]$, where $R^\star_-(\{h_{2n-1}\})$ is given by,

$$\begin{aligned}
R^\star_-(\{h_{2n-1}\}) &= [R^\star_-(\{h_{2n-3}\}) + G^\star_-(\{h_{2n-2}\})] \\
&- \sum_{m=0}^{z-1} \binom{z-1}{m} [R^\star_+(\{h_{2n-2}\}) + G^\star_+(\{h_{2n-1}\})]^m \\
&\times [R^\star_-(\{h_{2n-3}\}) + G^\star_-(\{h_{2n-2}\})]^{z-1-m} \\
&\times [p_{m+1}(h_{2n-1}) - p_{m+1}(h_{2n-2})],
\end{aligned} \quad (4.9)$$

and $G^\star_-(\{h_{2n}\})$ satisfies the self consistent equation

$$\begin{aligned}
G^\star_-(\{h_{2n}\}) &= \sum_{m=0}^{z-1} \binom{z-1}{m} [R^\star_+(\{h_{2n-2}\}) + G^\star_+(\{h_{2n-1}\})]^m \\
&\times [R^\star_-(\{h_{2n-1}\}) + G^\star_-(\{h_{2n}\})]^{z-1-m} \\
&\times [p_{m+1}(h_{2n-1}) - p_{m+1}(h_{2n})].
\end{aligned} \quad (4.10)$$

Therefore the decrease in magnetization, when the field is decreased from $h_{2n-1}$ to $h_{2n}$, is obtained from

$$\begin{aligned}
\text{Prob}&(s_O = -1; h_{2n} | s_O = +1; h_{2n-1}) \\
&= \sum_{m=0}^{z} \binom{z}{m} [R^\star_+(\{h_{2n-2}\}) + G^\star_+(\{h_{2n-1}\})]^m \\
&\quad \times [R^\star_-(\{h_{2n-1}\}) + G^\star_-(\{h_{2n}\})]^{z-m} [p_m(h_{2n-1}) - p_m(h_{2n})].
\end{aligned} \quad (4.11)$$

Similarly on the lower half, the increase in magnetization, when the field is increased from $h_{2n}$ to $h_{2n+1}$, is obtained from

$$\begin{aligned}
\text{Prob}&(s_O = +1; h_{2n+1} | s_O = -1; h_{2n}) \\
&= \sum_{m=0}^{z} \binom{z}{m} [R^\star_+(\{h_{2n}\}) + G^\star_+(\{h_{2n+1}\})]^m \\
&\quad \times [R^\star_-(\{h_{2n-1}\}) + G^\star_-(\{h_{2n}\})]^{z-m} [p_m(h_{2n+1}) - p_m(h_{2n})],
\end{aligned} \quad (4.12)$$

where $R^\star_+(\{h_{2n}\})$ is given by

$$\begin{aligned}
R^\star_+(\{h_{2n}\}) &= [R^\star_+(\{h_{2n-2}\}) + G^\star_+(\{h_{2n-1}\})] \\
&- \sum_{m=0}^{z-1} \binom{z-1}{m} [R^\star_+(\{h_{2n-2}\}) + G^\star_+(\{h_{2n-1}\})]^m \\
&\times [R^\star_-(\{h_{2n-1}\}) + G^\star_-(\{h_{2n}\})]^{z-1-m} \\
&\times [p_m(h_{2n-1}) - p_m(h_{2n})],
\end{aligned} \quad (4.13)$$

and $G_+^\star(\{h_{2n+1}\})$ is obtained by solving the self consistent equation

$$
\begin{aligned}
G_+^\star(\{h_{2n+1}\}) &= \sum_{m=0}^{z-1} \binom{z-1}{m} [R_+^\star(\{h_{2n}\}) + G_+^\star(\{h_{2n+1}\})]^m \\
&\quad \times [R_-^\star(\{h_{2n-1}\}) + G_-^\star(\{h_{2n}\})]^{z-1-m} \\
&\quad \times [p_m(h_{2n+1}) - p_m(h_{2n})].
\end{aligned} \quad (4.14)
$$

In Fig. 4.1, we have plotted first two minor hysteresis loops, generated by solving above equations, for three coordinated Bethe lattice. The random field distribution is given by Eq. (2.11) and we choose $\Delta = 1.5$.

## 4.2 Merging of different stable configurations

In the previous section, we see that the two ends of minor loop are at major loop, with the external field is differed by $2J$. In this section we generalize this to any stable configuration on any graphs. We prove that, for RFIM on a connected general graph $G$, for a given realization of random fields, all the stable configurations at external field $h$ go to unique stable configuration $C(h \pm 2z^\star J)$ when the field is monotonically changed to $h \pm 2z^\star J$; where $z^\star$ is the minimum number required such that, any connected subgraph $g$ of $G$ has at least one vertex such that, the number of edges in $g$ connected to that vertex is $\leq z^\star$. For example, $z^\star = 2$ for square lattice, and for Bethe lattice $z^\star = 1$.

*Proof.* — Consider two stable configurations $C_1(h)$ and $C_2(h)$ at field $h$. We can decompose vertices of the graph $G$, into sets: (1) $A_{uu}$, up-spins in both configurations, (2) $A_{ud}$, up-spins in $C_1$ and down-spins in $C_2$, (3) $A_{du}$, down-spins in $C_1$ and up-spins in $C_2$ and (4) $A_{dd}$, down-spins in both configurations. Consider when the external field is increased monotonically by $2z^\star J$. Since in the zero temperature dynamics, in the increasing field field the spins flip only once, and the order in which various spins are relaxed does not matter, we can increase the field in one step to $h + 2z^\star J$ and then first relax spins from sets $A_{ud}$ and $A_{du}$. Now the set $A_{ud}$ can be written as union of disjoint subsets $A_{ud}^{(1)}, A_{ud}^{(2)} \ldots$. Consider one such subset, which is on a subgraph $g$. On this subgraph, the local field at a vertex $i$ in configurations $C_1$ and $C_2$ at field $h$ are,

$$\ell_i^{(C_1(h))} = h_i + h + z_i^g J - z_i' J + f_i > 0, \quad (4.15)$$

$$\ell_i^{(C_2(h))} = h_i + h - z_i^g J + z_i' J + f_i < 0, \quad (4.16)$$

where $z_i^g$ are the number of vertices in $g$ connected to vertex $i$, $z_i'$ are the number of vertices in the set $A_{du}$ connected to $i$ and $f_i$ is the contribution to the local field from the sets $A_{uu}$ and $A_{dd}$. Since $\ell_i^{(C_1(h))} - \ell_i^{(C_2(h))} > 0$, from Eq. (4.15) and Eq. (4.16) we get $z_i^g - z_i' > 0$ or $z_i^g \geq 1$ as $z_i' \geq 0$. This means that, there can not be a subset of $A_{ud}$ which has only one element.

Now when the external field is increased to $h + 2z^\star J$, the local field at vertex $i$ in configuration $C_2$ becomes,

$$\ell_i' = \{h_i + h + z_i^\star J - z_i' J + f_i\} + \{(z_i^\star - z_i^g)J + 2z_i' J\}. \tag{4.17}$$

Therefore, at all those vertices in $g$ where $z_i^g \leq z^\star$ (there is at least one such vertex in $g$, by the definition of $z^\star$), the local field will become positive. So, the spins in $C_2$, at those vertices will flip up and the original subset will will shrink to a new one on a different reduced subgraph $g_\prime$ and the same argument holds for it also. Therefore, after iterative use of this relaxation procedure, all the subsets of $A_{ud}$ will become null sets and by the same argument, it is also true for $A_{du}$. Therefore, after relaxing all the spins from the sets $A_{ud}$ and $A_{du}$, the resultant unstable configurations $\widetilde{C_1}(h+2z^\star J)$ and $\widetilde{C_2}(h+2z^\star J)$ are identical, and hence relaxation of the remaining unstable spins from the set $A_{dd}$, will lead to the same final stable configuration. From the symmetry between up and down spins, it is obvious that the configuration $C_1(h)$ and $C_2(h)$ go to the same final configuration, when the field is decreased by $2z^\star J$.

For a square lattice $z^\star = 2$, as in any connected subgraph of it, there exists at least one vertex form which the number of edges connected to the subgraph is $\leq 2$. Therefore any two different stable configurations should merge to one configuration, when the field is increased or decreased by $4J$. In Fig. 4.3, we consider two stable spin configurations (a) and (b) at external field $h = 0$, on a square lattice, for a given realization of random fields. The lattice size is $50 \times 50$. The spins which are down in (a) and up in (b) are shown in (c) in black color. Now when we increase the external field to $4J$, configurations (a) and (b) evolve to new stable configurations (d) and (e) respectively. We see that these two configurations (d) and (e) are identical as seen from their difference configuration (c).

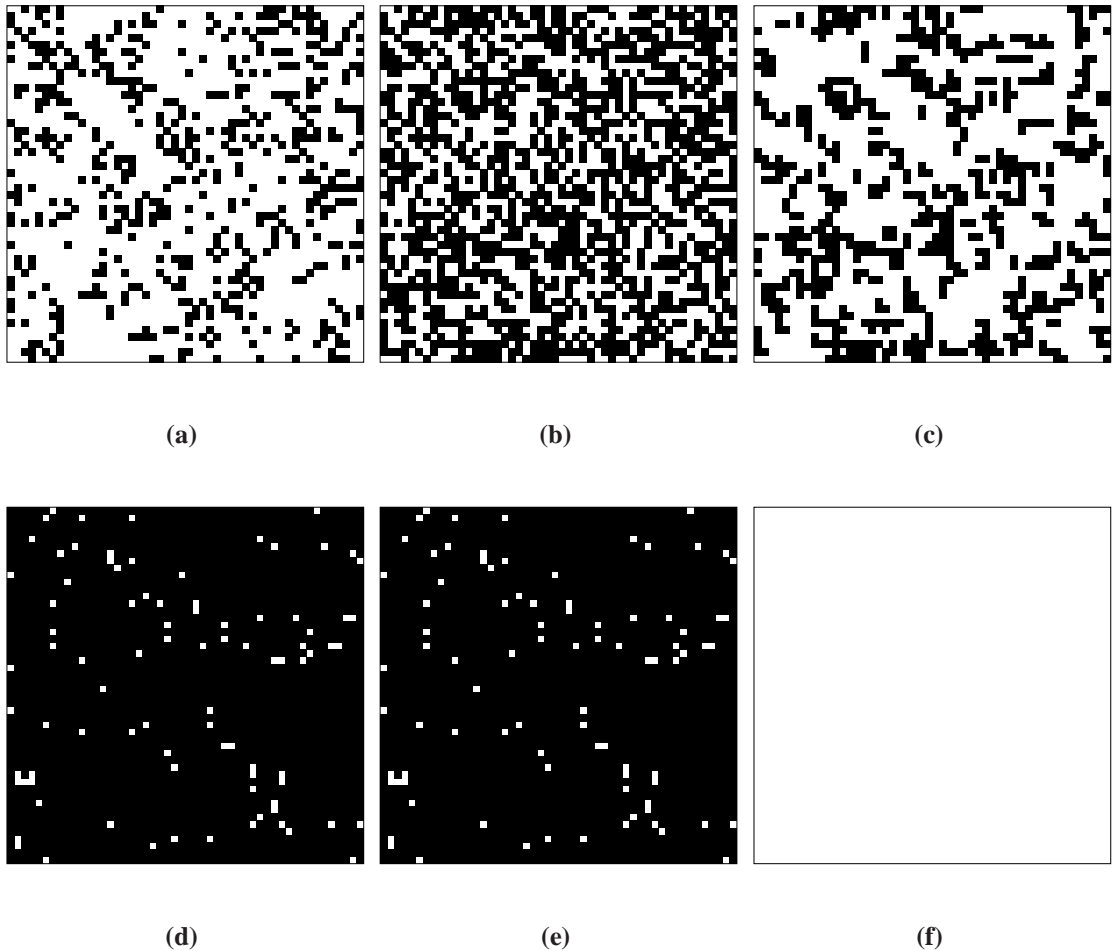

**Figure 4.3:** (a) and (b)are two different stable spin configurations at $h = 0$, with the same realization of random fields, on a square lattice of size $50 \times 50$. The up spins are represented by black and down spins by white color. (c) shows the difference between (a) and (b). The spins, which are down in (a) and up in (b) are represented by black color in (c). (d) and (e) are new stable configurations obtained from (a) and (b) respectively, at $h = 4J$. (f) shows the difference between (d) and (e).

*Chapter 5*

# Hysteresis on regular lattices in the low disorder limit

In general, Bethe approximation is expected to work well for noncritical properties. *Is the Bethe approximation is a good approximation for regular lattices?* This is the question we address in this chapter. Surprisingly, for asymmetrical distribution, the answer can be *no*.

In this chapter, we discuss the low disorder limit of the hysteresis loop in the random field Ising model (RFIM) on periodic lattices in two and three dimensions. We find that the behavior of hysteresis loops depends nontrivially on the coordination number $z$ (Sabhapandit et al. 2002). For $z = 3$, for continuous unbounded distributions of random fields, the hysteresis loops show no jump discontinuity of magnetization even in the limit of small disorder, but for higher $z$ they do. This is exactly as found in the exact solution on the Bethe lattice (Dhar et al. 1997).

As discussed in the introduction, random field Ising model was first studied in the context of possible destruction of long range order by arbitrarily weak quenched disorder in equilibrium systems. Accordingly the distribution of random field was assumed to be symmetrical. However, in hysteresis problem, the symmetry between up and down spins state is already broken by the specially prepared initial state (all down in our case), and the symmetry of the distribution plays no special role.

The analytical treatment of self-consistent equations on the Bethe lattice is immediately generalized to asymmetrical case. However, we find that for asymmetrical distributions the behavior of hysteresis loops in euclidean lattices can be quite different from that on the Bethe lattice. On hypercubical lattices in $d$ dimensions, there is an instability related to bootstrap percolation, that is absent on the Bethe lattice. This reduces the value of the coercive field $h_{\text{coer}}$ away from the Bethe lattice value $\mathcal{O}(J)$ to zero, where $J$ is the exchange coupling. We note that the limit $\Delta \to 0$ is somewhat subtle, as the system size $L^\star$ required for self-averaging diverges very fast for small $\Delta$, and the finite-size corrections to the thermodynamic limit tend to zero very slowly.

In the following, we shall assume that the distribution has a asymmetrical shape, given by

$$\phi(h_i) = \frac{1}{\Delta}\exp(-h_i/\Delta)\theta(h_i); \tag{5.1}$$

where $\theta$ is the step function. The mean value of $h_i$ can be made zero by a shift in the value of the external uniform field. Our treatment is easily extended to other continuous unimodal distributions. The exact form of $\phi(x)$ is not important, and other forms like $\exp(-x - e^{-x})$ which fall sharply for negative $x$ have the same behavior.

For a given distribution $\phi(h_i)$, we define $p_m(h)$ with $0 \leq m \leq z$ as the conditional probability that the local field at any site $i$ will be large enough so that it will flip up, if $m$ of its neighbors are up, when the uniform external field is $h$. Clearly

$$p_m(h) = \int_{(z-2m)J-h}^{\infty} \phi(h_i)\, dh_i. \tag{2.7}$$

Clearly, for any given value of $h$, the magnetization depends on the distribution $\phi(h_i)$ only through $p_m(h)$.

## 5.1 Hysteresis on three coordinated lattices

Consider first the case of the two-dimensional hexagonal lattice with $z = 3$. For periodic boundary conditions, if $\Delta = 0$, starting with a configuration with all spins down, clearly one has $h_{\text{coer}} = 3J$. For $\Delta \neq 0$, the site with the largest local field flips first, and then if $h > J$, $p_1(h) = 1$, this causes neighbors of the flipped spin to flip, and their neighbors, and so on. Thus, so long as there is at least one flipped spin, all other spins also flip, and the magnetization is $1$. The largest local field in a system of $L^2$ spins is of order $2\Delta \ln L$. Once this spin turns up, other spin will flip also up, causing a jump in magnetization from a value $\approx -1$ to a value $+1$ in each sample. Hence the coercive field, (the value of $h$ where magnetization changes sign) *to lowest order in* $\Delta$, is given by

$$h_{\text{coer}} = 3J - 2\Delta \ln L, \quad \text{for} \quad 1 \ll \ln L \ll J/\Delta. \tag{5.2}$$

Sample to sample fluctuations in the position of the jump are of order $\Delta$. On averaging over disorder, the magnetization will become a smooth function of $h$, with the width of the transition region being of order $\Delta$.

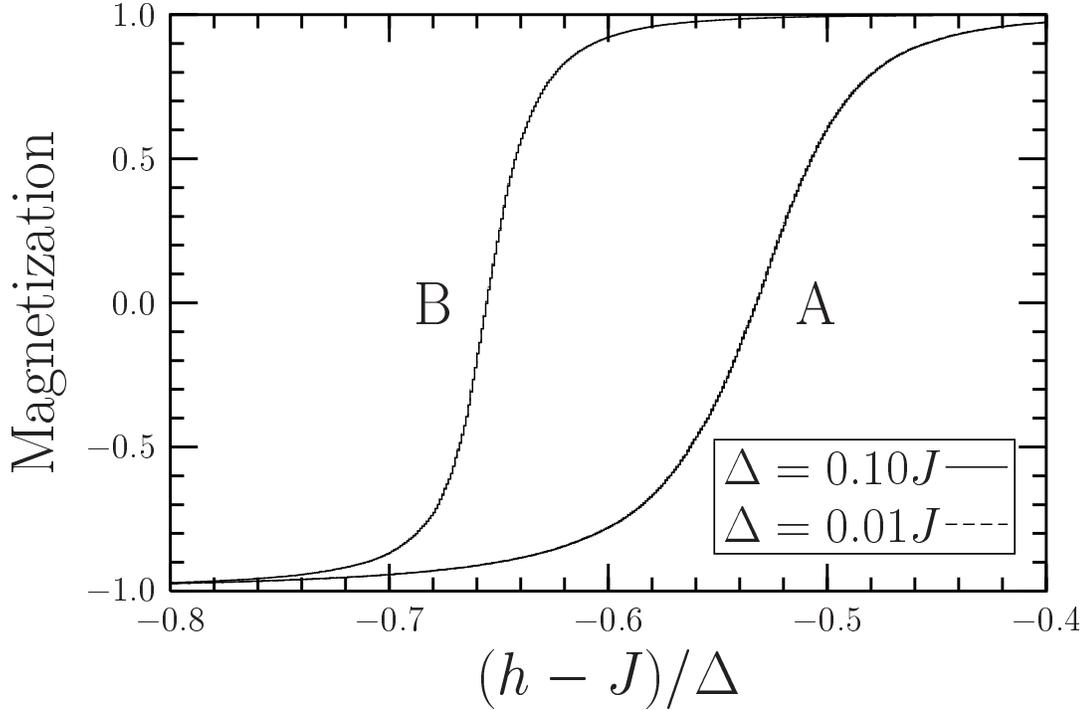

**Figure 5.1:** Magnetization in the increasing field. The curves for the two values of $\Delta$ coincide. Curves A is for hexagonal lattice of size $4096^2$ and B is for a three coordinated lattice in three dimensions [see Fig. 5.2] of size $256^3$.

For a fixed $\Delta \ll J$, if $L$ is increased to a value near $\exp(J/\Delta) \equiv L^{\star}_{\text{hex}}$, $h_{\text{coer}}$ decreases to a value near $J$. For $h \approx J$, $p_1(h)$ is no longer nearly 1, but $p_0(h) \simeq 0$, and $p_2(h) \simeq p_3(h) \simeq 1$. The value of magnetization depends only on $p_1(h)$, which is a function of $\widetilde{h} = (h - J)/\Delta$. As $\widetilde{h}$ increased from $-\infty$, $p_1(h)$ increases continuously from 0 to 1.

Note that for $\Delta = 0.01J$, $L^{\star}_{\text{hex}} \sim 10^{43}$. Therefore it is impossible to study the large $L$ limit with the available computers. To avoid the problem of probability of nucleation being very small for $h$ near $J$, we made the local field at a small fraction of randomly chosen sites very large, so that these spins are up at any $h$. The number of such spins we choose to be of order $L$, so that their effect on the average magnetization is negligible. Introduction of these "nucleation centers" makes $L^{\star} \approx \mathcal{O}(\sqrt{L})$ ( the average separation between centers), and $h_{\text{coer}}$ drops to a value near $J$, so that, we can study the large $L$ limit with available computers. For $L > L^{\star}_{\text{hex}}$, the behavior of hysteresis loops becomes independent of $L$.

In Fig. 5.1, curve A shows the result of a simulation on the hexagonal lattice with $L = 4096$, and periodic boundary condition. We see that magnetization no longer under-goes a single large jump, but many small jumps. In the figure, we also show the plot of magnetization when the random field at each site is decreased by a factor 10. This changes the value $\Delta$ from $0.1J$ to $0.01J$. However, plotted as a function of $\tilde{h}$, the magnetization for these two different values (for small $\Delta$) fall on top of each other *for the same realization of disorder* (except for the overall scale $\Delta$). Thus we can decrease $\Delta$ further to arbitrarily small values, and the limit of $\Delta \to 0$ is straightforward for each realization of disorder. Then, averaging over disorder, for a fixed $\Delta$, we see that $h_{\text{coer}}$ tends to the value $J$ as $\Delta$ tends to $0$. Also, we see that there is no macroscopic jump-discontinuity for any non-zero $\Delta$.

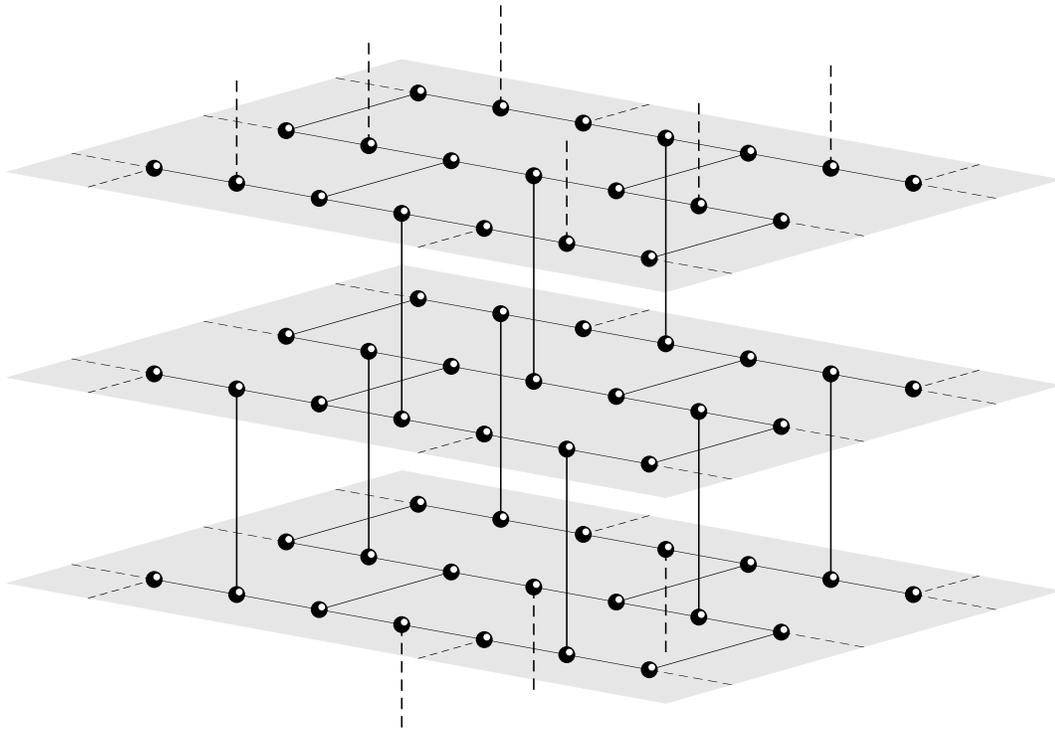

**Figure 5.2:** A three coordinated lattice ($z = 3$) in three dimensions.

We also show in Fig. 5.1 [curve B], the results of simulation of a 3-dimensional lattice with $z = 3$ [shown in Fig. 5.2] of size $256^3$ with periodic boundary condition. The behavior is qualitatively same as that in two dimensions. The value of $h_{\text{coer}} = J$ in the limit $\Delta \to 0$ is same for symmetrical distribution, and also is the same as predicted by the Bethe approximation.

## 5.2 Bootstrap instability in RFIM on square lattice

On the square lattice also, the value of $h_{\text{coer}}$ is determined by the need to create a nucleation event. Arguing as before, we see that $h_{\text{coer}}$ to lowest order in $\Delta$ is given by

$$h_{\text{coer}} \approx 4J - 2\Delta \ln L, \qquad \text{for} \quad 1 \ll \ln L \ll J/\Delta. \tag{5.3}$$

Adding a small number of nucleation sites suppresses this slow transient, and lowers $h_{\text{coer}}$ from $4J$ to a value near $2J$. However, in this case, even after adding the nucleation centers, the system shows a large single jump in magnetization, indicating the existence of another instability. We observed in the simulation that at low $\Delta$, as $h$ is increased, the domains of up spins grow in rectangular clusters [see Fig. 5.3] and at a critical value of $h_{\text{coer}}$, one of them suddenly fills the entire lattice. This value $h_{\text{coer}}$ fluctuates a bit from sample to sample.

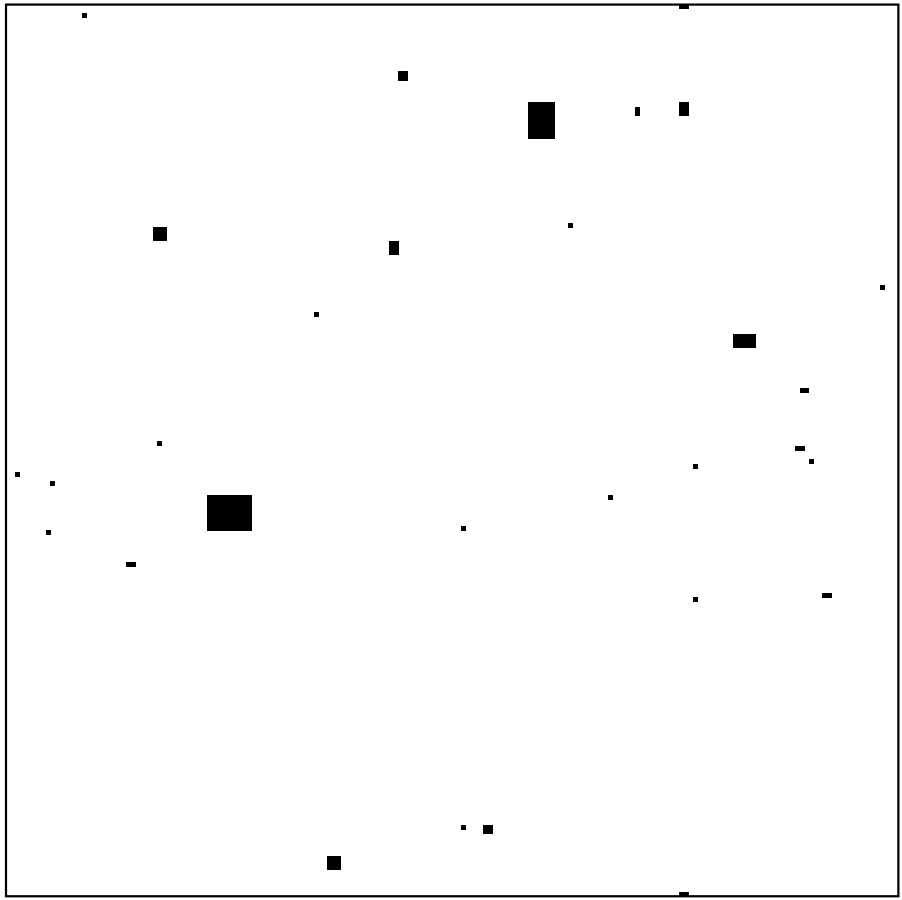

**Figure 5.3:** A snapshot of the up-spins just before the jump ($h = 1.998243 J$). The lattice size is $200 \times 200$ and $\Delta = 0.001 J$. Initial configuration is prepared with $0.05\%$ up-spins.

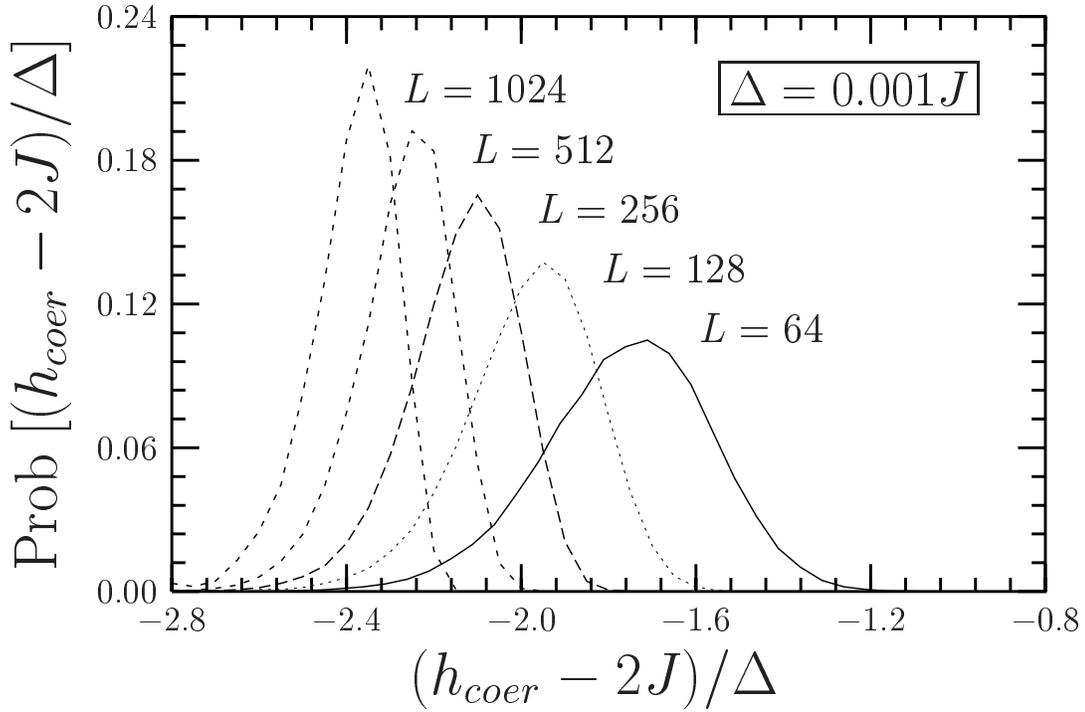

**Figure 5.4:** Distribution of the scaled coercive field on a square lattice for different lattice size $L^2$.

In Fig. 5.4 we have plotted the distribution of the scaled variable $\widetilde{h_c} = (h_{coer} - 2J)/\Delta$ for different system sizes $L$, for $\Delta = 0.001J$. The number of different realizations varies from $10^4$ (for the largest $L$) to $10^5$ (for the smallest $L$). Note that the distribution shifts to the left with the increasing system size, and becomes narrower.

This instability can be understood in terms of bootstrap percolation process $BP_m$ (see Adler 1991, for a review). Bootstrap percolation was first considered by Chalupa et al. (1981) (also Kogut and Leath 1981) and was subsequently studied by many others in a variety of contexts. The process $BP_m$ is define as follows: On a $d-$dimensional lattice, sites are independently occupied with a probability $p$ and the resulting configuration is taken as the initial configuration, which is evolved by the following rules:

 (a) the occupied sites remain occupied forever,
 (b) an unoccupied site having at least $m$ occupied neighbors, becomes occupied.

For $m = 2$, on a square lattice, in the final configuration, the sites which are occupied form disjoint rectangles, like the cluster of up-spins in Fig. 5.3. It has been proved that in the thermodynamic limit of large $L$, for any initial concentration $p > 0$, in the final configuration all sites are occupied with probability 1 (Aizenman and Lebowitz 1988).

In the random field Ising model on a square lattice, for the asymmetric distribution

[Eq. 5.1] for $h > 0$, $p_m = 1$ for $m \geq 2$, and any spins with more than one up-neighbors flips up. Therefore, stable clusters of up spins are rectangular in shape. The growth of domains of up spins is same as in the bootstrap percolation process $BP_2$.

Consider a rectangular cluster of up spins, of length $l$ and width $m$. Let $P(l, m)$ be the probability that, if this rectangle is put in a randomly prepared background of density $p_1(h)$, this rectangle will grow by the $BP_2$ process to fill the entire space. The probability that the random fields at any sites neighboring this rectangle will be large enough to cause it to flip up is $p_1(h)$. The probability that there is at least one such site along each of two adjacent sides of length $l$ and $m$ of the rectangle is $(1 - q^l)(1 - q^m)$, where $q = 1 - p_1(h)$. Once these spins flip up, this induces all the other spins along the boundary side to flip up and the size of the rectangle grows to $(l+1) \times (m+1)$. Therefore

$$P(l, m) \geq (1 - q^l)(1 - q^m) P(l+1, m+1). \tag{5.4}$$

Thus the probability of occurrence of a nucleation which finally grows to fill the entire lattice is

$$\begin{aligned} P_{\text{nuc}} &\geq p_0(h) \prod_{j=1}^{\infty} (1 - q^j)^2 \\ &\approx p_0(h) \prod_{j=1}^{\infty} [1 - \exp\{-p_1(h)j\}]^2 \\ &\approx p_0(h) \exp\left(-\frac{\pi^2}{3 p_1(h)}\right), \quad \text{for small } p_1(h). \end{aligned} \tag{5.5}$$

The condition to determine $h_{\text{coer}}$ is that for this value of $h$, $P_{\text{nuc}}$ becomes of order $1/L^2$, so that we get

$$p_0(h_{\text{coer}}) \exp\left(-\frac{\pi^2}{3 p_1(h_{\text{coer}})}\right) \approx \frac{1}{L^2}. \tag{5.6}$$

This equation can be solved for $h_{\text{coer}}$ for any given $L$. For the distribution given by Eq. (5.1), this becomes

$$\exp\left(\frac{h_{\text{coer}} - 4J}{\Delta}\right) \exp\left[-\frac{2\pi^2}{3} \exp\left(\frac{-h_{\text{coer}} + 2J}{\Delta}\right)\right] \approx \frac{1}{L^2}, \quad \text{for } h_{\text{coer}} < 2J. \tag{5.7}$$

Therefore, the leading $L$-dependence of $h_{\text{coer}}$, to lowest order in $\Delta$ is given by

$$h_{\text{coer}} \approx 2J - \Delta \ln\left[\frac{3}{\pi^2}(\ln L - J/\Delta)\right], \quad \text{for} \quad J/\Delta \ll \ln L \ll \exp(2J/\Delta). \tag{5.8}$$

This agrees with our observation that the scaled critical field $\widetilde{h}_c$ shifts to the left with increasing system size. The width of the distribution of over which the coercive field varies, can be calculated from the width over which the probability of having at least one nucleation in the entire lattice, i.e. $1 - (1 - P_{\text{nuc}})^{L^2} \approx 1 - \exp(-P_{\text{nuc}} L^2)$, changes from almost zero to almost unity:

$$\delta h_{\text{coer}} \sim \frac{\Delta}{\ln L} \tag{5.9}$$

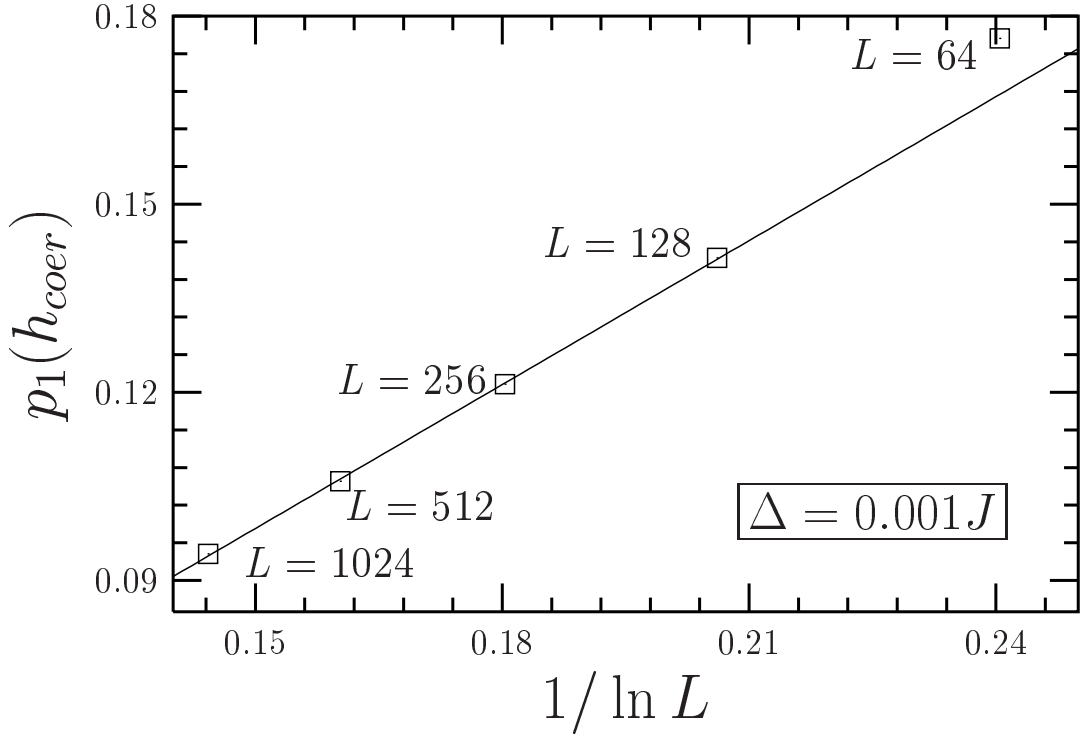

**Figure 5.5:** $p_1(h_{\text{coer}})$ vs. $1/\ln L$ for square lattice.

Therefore, for any fixed $\Delta > 0$, the jump will smeared out on averaging over disorder. Only in the limit $\Delta \to 0$ and $L \to \infty$, the average magnetization will show a jump discontinuity.

To test the validity of Eq. (5.6) in simulations, we put $p_0(h) = 0.005$ independent of $h$. Eq. (5.6) then simplifies to

$$p_1(h_{\text{coer}}) \approx \frac{\pi^2}{6 \ln L}. \tag{5.10}$$

In Fig. 5.5, we have plotted $p_1$ for the mean $h_{\text{coer}}$ from Fig. 5.4 versus $1/\ln L$. The graph is approximately a straight line, which agrees with Eq. (5.10). The slope of the line is $0.765 \pm 0.009$, less than in Eq. (5.10), which only gives an upper bound to $h_{\text{coer}}$.

If $h > 0$, we will have $p_2 = 1$, and bootstrapping ensures that so long as $p_0 > 0$, we will have all spins up in the limit of large $L$. This implies that $h_{\text{coer}} = 0$ in this limit.

If there are sites with large negative quenched fields, the bootstrap growth stops at such sites. Hence the bootstrap instability cannot be seen for symmetric distributions. For a symmetrical distribution of random fields, the average distance from a nucleus to the nearest spin, which does not flips up even if it has two up neighbors, is $L^\star = 1/\sqrt{1 - p_2(h)}$. Therefore the average area covered by a nucleus is

$$L^2 \prod_{j=1}^{L} (1 - q^j)^2 \quad \text{for} \quad L < L^\star \tag{5.11a}$$

and

$$L^{\star 2} \prod_{j=1}^{L^\star} (1 - q^j)^2 \quad \text{for} \quad L > L^\star. \tag{5.11b}$$

The condition to determine $h_{\text{coer}}$ is that for this value of $h$, the average area due to the growth becomes $\mathcal{O}(L^2)$, i.e.

$$p_0(h_{\text{coer}}) L^2 \times \text{average area covered by a nucleus} \approx \mathcal{O}(L^2). \tag{5.12}$$

From Eq. (5.11b) and Eq. (5.12), in the limit $L \to \infty$,

$$\frac{p_0(h_{\text{coer}})}{1 - p_2(h_{\text{coer}})} \prod_{j=1}^{L^\star} (1 - q^j)^2 \approx \mathcal{O}(1). \tag{5.13}$$

Now $p_0(2J) = 1 - p_2(2J)$, and the product $\prod_{j=1}^{L^\star}(1-q^j)^2$ is $\mathcal{O}(1)$ at $h = 2J$ as $p_1(2J) = 1/2$. Therefore, $h_{\text{coer}} \approx \mathcal{O}(2J)$.

Even if the quenched fields are only positive, the instability does not occur on lattices with $z = 3$. On such lattices, if the unoccupied sites percolate, there are infinitely extended lines of unoccupied sites in the lattice. These cannot not become occupied by bootstrapping under $BP_2$. Thus the critical threshold for $BP_2$ on such lattices is not $0$.

## 5.3 Bootstrap instability in RFIM on cubic lattice

The arguments for large void instability can be easily extended to higher dimensions. In $d = 3$, if $h > 0$, then $p_m(h) = 1$ for $m \geq 3$, therefore the spin flip process is similar to the spanning process of three dimensional $BP_3$ (Cerf and Cirillo 1999). In this case, it is known that for any initial non-zero density, in the thermodynamical limit, the final configuration has all sites occupied with probability $1$. The clusters of up-spins grow as cuboids, and at each surface of the cluster, the nucleation process is similar to that in two dimension. Let $\epsilon$ be the probability that, a nucleation occurs at a given point of a surface of the clusters of up spins which sweeps the entire two dimensional plane at $h$.

$$\epsilon \approx p_1(h) \exp\left(-\frac{\pi^2}{3p_2(h)}\right). \tag{5.14}$$

The probability that, there exist at least one nucleation which sweeps the entire plane of size $l \times l$, is $1 - (1-\epsilon)^{l^2}$. Therefore, the probability $P_{\text{nuc}}$, that a nucleation sweeps the entire three dimensional lattice at $h$ satisfies

$$\begin{aligned} P_{\text{nuc}} &\geq p_0(h) \prod_{l=1}^{\infty} \left[1 - (1-\epsilon)^{l^2}\right]^3 \\ &\approx p_0(h) \prod_{l=1}^{\infty} \left[1 - \exp(-\epsilon l^2)\right]^3 \\ &\approx \exp(-A/\sqrt{\epsilon}), \quad \text{for small } \epsilon; \end{aligned} \tag{5.15}$$

where $A = \frac{3}{2}\sqrt{\pi}\zeta(3/2)$. $h_{\text{coer}}$ is determined by the condition that $P_{\text{nuc}}$ must be of the order $1/L^3$:

$$p_0(h_{\text{coer}}) \exp\left[-\frac{A}{\sqrt{p_1(h_{\text{coer}})}} \exp\left(\frac{\pi^2}{6p_2(h_{\text{coer}})}\right)\right] \approx \frac{1}{L^3}. \tag{5.16}$$

The leading $L$-dependence of $h_{\text{coer}}$ is different in different ranges of $h_{\text{coer}}$, depending on whether the strongest dependence of the left-hand side comes from variation of $p_0(h), p_1(h)$ or $p_2(h)$.

In the range $4J < h_{\text{coer}} < 6J$: $p_m = 1$, for $m \geq 1$. Then we must have $p_0(h_{\text{coer}}) \sim 1/L^3$, which for the distribution given by Eq. (5.1) results

$$h_{\text{coer}} \approx 6J - 3\Delta \ln L. \tag{5.17a}$$

The corresponding range of $L$, for the validity of of above equation is $1 \ll \ln L \ll (2J/3\Delta)$.

In the range $2J < h_{\text{coer}} < 4J$: $p_m = 1$, for $m \geq 2$. Then in Eq. (5.16) the left hand side varies as $\exp\left[-A'/\sqrt{p_1(h_{\text{coer}})}\right]$, which gives

$$h_{\text{coer}} \approx 4J - 2\Delta \ln\left(\ln L - \frac{2J}{3\Delta}\right), \tag{5.17b}$$

which is valid in the range $(2J/3\Delta) \ll \ln L \ll \exp(2J/\Delta)$.

In the range $0 < h_{\text{coer}} < 2J$: $p_m = 1$ for $m \geq 3$. Then from Eq. (5.16), to the lowest order in $\Delta$, we get

$$h_{\text{coer}} \approx 2J - \Delta \ln\ln\left(\ln L - \frac{2J}{3\Delta}\right), \tag{5.17c}$$

for $\exp(2J/\Delta) \ll \ln L \ll \exp(\exp(2J/\Delta))$.

In the limit $L \gg L^\star_{\text{cub}} = \exp(\exp(\exp(2J/\Delta)))$, the loop becomes independent of $L$, with $h_{\text{coer}} \to 0$. We have also verified the existence of jump in numerical simulation for $z = 4$ (diamond lattice) in three dimensions.

# Chapter 6

# Discussion

Analytical treatment of problems having quenched disorder is usually difficult. There are few models having nontrivial quenched disorder that can be solved exactly. In this thesis, we set up exact self-consistent equations for the avalanche distribution function for the RFIM on a Bethe lattice. We were able to solve these equations explicitly for the rectangular distribution of the quenched field, for the linear chain $z = 2$, and the 3-coordinated Bethe lattice. For more general coordination numbers, and general continuous distributions of random fields, we argued that for very large disorder, the avalanche distribution is exponentially damped, but for small disorder, generically, one gets a jump in magnetization, accompanied by a square-root singularity. For field-strengths just below corresponding to the jump discontinuity, we showed that the avalanche distribution function has a power-law tail of the form $s^{-3/2}$. The integrated avalanche distribution then varies as $s^{-5/2}$ for large $s$.

We have also studied the behavior the return loop, when the external field is increased from $-\infty$ to some value $h_1$, and then decreased to a lower value $h_2$ and again increased to the previous extremum value $h_1$. We set up exact self-consistent equations to determine the magnetizations on all minor loops for arbitrary distributions of random fields.

Some unexpected features of the solution deserve mention. Firstly, we find that the behavior of the self-consistent equations for $z = 3$ is qualitatively different from that for $z > 3$. The behavior for the linear chain ($z = 2$) is, of course, expected to be different from higher $z$. One usually finds same behavior for all $z > 2$. Mathematically, the reason for this unusual dependence is that the mechanism of two real solutions of the polynomial equation merging, and both becoming unphysical (complex) is not available for $z = 3$. Here the self-consistency equation is a quadratic, and from physical arguments, at least one of the roots must be real. That a Bethe lattice may show non-generic behavior for low coordination numbers has been noted earlier by Ananikyan et al. (1994) in their study of the Blume-Emery-Griffiths model on a Bethe lattice. These authors observed that the qualitative behavior for $z < 6$ is different from that for $z \geq 6$. To find out whether this unusual $z$ dependence of the hysteresis loop persists on regular lattices, we study it on the regular lattices in two and three dimension, in the limit of low disorder. We find that for asymmetrical distributions of random fields, there is a instability which is not present in three coordinated lattices, and hence the hysteresis curve is continuous for such lattices.

The second point we want to emphasize is that here we find that the power-law tail in the distribution function is accompanied by the first-order jump in magnetization. Usually, one thinks of critical behavior and first-order transitions as mutually exclusive, as first-order jump pre-empts a build-up of long-ranged correlations, and all correlations remain finite-ranged across a first-order transition. This is clearly not the case here. In fact, the power-law tail in the avalanche distribution disappears, when the jump disappears. A similar situation occurs in equilibrium statistical mechanics in the case of a Heisenberg ferromagnet below the critical temperature. As the external field $h$ is varied across zero, the magnetization shows a jump discontinuity, but in addition has a cusp singularity for small fields[†]. But in this case the power-law tail is seen on *both sides of the transition*.

Note that for most values of disorder, and the external field, the avalanche distribution is exponentially damped. We get robust power law tails in the distribution, only if we integrate the distribution over the hysteresis cycle across the magnetization jump. But, in this case, the control parameter $h$ is swept across a range of values, in particular across a (non-equilibrium) phase transition point! In this sense, while no explicit fine-tuning is involved in an experimental setup, this is not a self-organized critical system in the usual sense of the word. Recently Pázmándi et al. (1999) have argued that the hysteretic response of the Sherrington-Kirkpatrick model to external fields at zero temperature shows self-organized criticality for all values of the field. However, this seems to be because of the presence of infinite-ranged interactions in that model.

In chapter 3, we discussed the behavior of avalanche distribution for various distributions of random fields. A general question concerns the behavior of the avalanches for more general probability distributions. Clearly, if $p(h_i)$ has a discrete part, it would give rise to jumps in $p_i$ as a function of $h$, and hence give rise to several jumps in the hysteresis loop. These could preempt the cusp singularity mechanism which is responsible for the

---

[†]Below $T_c$, the magnetization goes as,

$$m \sim sign(h) \left[ m_0 + A|h|^{(d-2)/2} \right], \quad \text{as} \quad h \to 0, \quad 2 < d < 4 \quad \text{(see Parisi 1988)}.$$

power-law tails. If the distribution $p(h_i)$ is continuous, but multimodal, then it is possible to have more than one first order jump in the magnetization[†]. This is confirmed by explicit calculation in some simple cases. If $p(h_i)$ has power-law singularities, these would also lead to power-law singularities in $p_i$, and hence in $P^\star(h)$. Even for purely continuous distributions, the merging of two roots as the magnetic field varies need not always occur. For example, it is easy to check that for the rectangular distribution, even for $z \geq 4$, we do not get a power law tail for any value of $\Delta$. The precise conditions necessary for the occurrence of the power-law tail needs to be investigated further.

Finally, we would like to mention some other open questions. Our analysis relied heavily on the fact that initial state was all spins down. Of course, we can start with other initial conditions. For example, start with the equilibrium state at temperature $T_0$ and field $h_0$, and then quench to zero temperature. Our present treatment cannot be applied to these cases as finite temperature brings about very nontrivial coorrelations between spins. It would be interesting to set up self-consistent field equations for them. In case of minor loops also, we have always started with $h = -\infty$ and then vary the field cyclically. Moreover, to find the magnetization at some particular point of the hysteresis curve, we start with the previous extremal field and change the field to the new value in one jump and then relax the system. It would be useful to find out some dynamical relations by which system can be evolved from any state by changing the field in infinitesimal steps.

Another extension would be to make the rate of field-sweep comparable to the single-spin flip rate (still assuming zero temperature dynamics). This would mean some large avalanches in different parts of the sample could be evolving simultaneously. Then one could study the sweep-rate dependence of the hysteresis loops, and the frequency dependence of the Barkhausen noise spectra. This is perhaps of some relevance in real experimental data, and would also make contact with other treatments of Barkhausen noise that focus on the domain wall motion.

Another case of some interest is other type of disorder e.g. the site-dilution case discussed by Tadić (1996). It seems plausible from the structural stability of the mechanism which leads to the cusp singularity just before the jump-discontinuity in magnetization, in our model, introduction of site dilution would not change the qualitative behavior of solutions.

We hope that many of these issues will be resolved in the next few years.

---

[†]This would happen if $P^\star(h)$ as a function of $h$ shows a 'double S' curve. Then there must be at least $4$ values of $h$ for which the slope of the curve is infinite. This is possible only if the equation determining $P^\star_{disc}$ [variant of Eq. (3.27)] is at least a quartic, hence only if $z \geq 6$.

# Appendix A

## A.1 Avalanche distribution on a linear chain

For the case of a linear chain, the self-consistent equation, for the probability $P^\star$ [Eq. (2.9)] is a linear equation, whose solution is,

$$P^\star(h) = \frac{p_0}{[1 - (p_1 - p_0)]}. \tag{A.1}$$

For $h < 2J - \Delta$, $p_0$ is zero, and hence $P^\star(h)$ is zero, and all the spin remain down (region A in Fig. 3.2).

For $h > 2J - \Delta$, and $\Delta < J$, $p_1$ is 1 whenever $p_0$ is nonzero. Then from Eq. A.1, $P^\star(h)$ becomes 1. Thus, for $\Delta < J$, we get a rectangular loop and the system changes from all spins down to all spins up state in a single big avalanche.

For $\Delta > J$, $p_1 - p_0$ equals $J/\Delta$ and is independent of $h$, in the range $2J - \Delta < h < \Delta$. Thus $P^\star(h)$ is a linear function of $h$ in this range, increasing from 0 to 1.

Defining

$$\epsilon = \frac{1}{2}\left(1 + \frac{h}{\Delta} - \frac{2J}{\Delta}\right), \tag{A.2}$$

we obtain the expression for $P^\star$ as

$$P^\star(h) = \begin{cases} 0 & \text{for } \epsilon < 0, \\ \frac{\epsilon}{1 - J/\Delta} & \text{for } 0 \leq \epsilon \leq 1 - J/\Delta, \\ 1 & \text{for } \epsilon > 1 - J/\Delta. \end{cases} \tag{A.3}$$

Using Eq. (3.6), the expression for $Q_0$ is,

$$Q_0 = (1 - p_1) - (p_2 - p_1)P^\star(h). \tag{A.4}$$

The generating function $Q(x)$ obtained from the self-consistent equation [Eq. (3.9)] is,

$$Q(x) = \frac{Q_0 + xP^\star(p_2 - p_1)}{1 - x(p_1 - p_0)}, \tag{A.5}$$

and the generating function $G(x|h)$ given by Eq.( 3.12) becomes ,

$$G(x|h) = x\left\{[Q(x)]^2\phi(2J - h) + 2P^\star[Q(x)]\phi(-h) + P^{\star 2}\phi(-2J - h)\right\}. \quad (A.6)$$

Now if $\Delta > 2J$, and $-\Delta + 2J < h < \Delta - 2J$ (region B in Fig. 3.2),

$$(p_2 - p_1) = (p_1 - p_0) = J/\Delta,$$
$$\phi(2J - mJ - h) = \frac{1}{2\Delta} \quad \text{for} \quad m = 0, 1, 2;$$
$$\text{and} \quad P^\star + Q_o = 1 - J/\Delta.$$

Thus

$$Q(x) = \frac{1 - (J/\Delta)}{1 - (J/\Delta)x} - P^\star, \quad (A.7)$$

and

$$G(x|h) = \frac{x}{2\Delta}[P^\star + Q(x)]^2 = \frac{x}{2\Delta}\frac{(1 - J/\Delta)^2}{(1 - xJ/\Delta)^2}. \quad (A.8)$$

Expanding $G(x|h)$ in powers of $x$, we get the probability distribution of avalanches in region B given by Eq. (3.14) of sec. 3.2.1.

In the region C, $p_2$ saturates to value 1, $\phi(-2J-h)$ becomes zero and $(p_2-p_1)$ becomes $(1 - J/\Delta - \epsilon)$. Thus we get,

$$Q_0 = \frac{(1 - J/\Delta - \epsilon)^2}{(1 - J/\Delta)}. \quad (A.9)$$

In terms of $P^\star$ and $Q_0$ we get

$$Q(x) = \frac{Q_0 + xP^\star[1 - 2(J/\Delta) - \epsilon]}{1 - (J/\Delta)x}, \quad (A.10)$$

and

$$G(x|h) = \frac{x}{2\Delta}\left\{[P^\star + Q(x)]^2 - P^{\star 2}\right\}. \quad (A.11)$$

Expanding $G(x|h)$ in powers of $x$ we get , in region C

$$G_1(h) = \frac{1}{2\Delta}\left[(P^\star + Q_0)^2 - P^{\star 2}\right], \quad (A.12)$$

and

$$G_s(h) = [A_1's + A_2']\left(\frac{J}{\Delta}\right)^s, \quad \text{for} \quad s \geq 2. \quad (A.13)$$

Here $A_2$ and $B_2$ have no dependence on $s$ but are explicit functions of $h$

$$A_1' = \frac{1}{2\Delta}\left[\frac{1}{(J/\Delta)}(P^\star + Q_0)^2 + \frac{1}{(J/\Delta)^2}\left(1 - \frac{2J}{\Delta} - \frac{h}{\Delta}\right)(P^\star + Q_0)P^\star \right.$$
$$\left. + \frac{1}{4(J/\Delta)^3}\left(1 - \frac{2J}{\Delta} - \frac{h}{\Delta}\right)^2 P^{\star 2}\right],$$

$$A_2' = \frac{-1}{\Delta}\left[\frac{1}{2(J/\Delta)^2}\left(1 - \frac{2J}{\Delta} - \frac{h}{\Delta}\right)(P^\star + Q_0)P^\star + \frac{1}{4(J/\Delta)^3}\left(1 - \frac{2J}{\Delta} - \frac{h}{\Delta}\right)^2 P^{\star 2}\right].$$

Integrating over $h$ from $-\infty$ to $\infty$ we get the integrated avalanche distribution $D_s$,

$$D_s = \int_{-\infty}^{\infty} G_s(h)dh, \tag{A.14}$$

where

$$D_1 = \frac{1}{(1-J/\Delta)^2}\left[1 - 6\left(\frac{J}{\Delta}\right) + 14\left(\frac{J}{\Delta}\right)^2 - \frac{46}{3}\left(\frac{J}{\Delta}\right)^3 + \frac{47}{6}\left(\frac{J}{\Delta}\right)^4 - \frac{9}{5}\left(\frac{J}{\Delta}\right)^5\right], \tag{A.15}$$

and, for $s \geq 2$,

$$D_s = (A_2 s + B_2)\left(\frac{J}{\Delta}\right)^s, \tag{A.16}$$

with

$$A_2 = \frac{1}{30(J/\Delta)}\left[30 - 110\left(\frac{J}{\Delta}\right) + 135\left(\frac{J}{\Delta}\right)^2 - 54\left(\frac{J}{\Delta}\right)^3\right],$$

$$B_2 = \frac{1}{15(1-J/\Delta)}\left[5 - 10\left(\frac{J}{\Delta}\right) + 4\left(\frac{J}{\Delta}\right)^2\right].$$

## A.2 Avalanche distribution on a three coordinated Bethe lattice

For $z = 3$, the self-consistent equation for $P^\star(h)$ [Eq. (2.9)] is a quadratic equation,

$$[(p_2 - p_1) - (p_1 - p_0)]P^\star(h)^2 + [2(p_1 - p_0) - 1]P^\star(h) + p_0 = 0. \tag{A.17}$$

For the rectangular distribution, the coefficient of $P^{\star 2}$ is zero for a range of $h$-values, and $P^\star(h)$ is still a piece wise linear function of $h$

$$P^\star(h) = \begin{cases} 0 & \text{for } \epsilon < 0, \\ \frac{\epsilon}{1-2(J/\Delta)} & \text{for } 0 \leq \epsilon \leq 1 - 2(J/\Delta), \\ 1 & \text{for } \epsilon > 1 - 2(J/\Delta), \end{cases} \tag{A.18}$$

where $\epsilon$ is defined as,

$$\epsilon = \frac{1}{2}\left(1 + \frac{h}{\Delta} - \frac{3J}{\Delta}\right). \tag{A.19}$$

The self-consistent equation for $Q(x)$ [Eq. (3.9)] becomes,

$$x(p_1 - p_0)[Q(x)]^2 + [2xP^\star(p_2 - p_1) - 1]Q(x) + xP^{\star 2}(p_3 - p_2) + Q_0 = 0, \tag{A.20}$$

where $Q_0$ is obtained [Eq. (3.6)] as

$$Q_0 = (1 - p_1) - 2(p_2 - p_1)P^\star + [(p_2 - p_1) - (p_3 - p_2)]P^{\star 2}, \tag{A.21}$$

and the expression for $G(x|h)$ [Eq. (3.12)] becomes,

$$G(x|h) = x\left\{[Q(x)]^3\phi(3J-h) + 3[Q(x)]^2 P^\star\phi(J-h)\right.$$
$$\left. + 3[Q(x)]P^{\star 2}\phi(-J-h) + P^{\star 3}\phi(-3J-h)\right\}. \qquad (A.22)$$

Now in the region B,

$$(p_3 - p_2) = (p_2 - p_1) = (p_1 - p_0) = J/\Delta,$$
$$\phi(3J - 2mJ - h) = \frac{1}{2\Delta} \quad \text{for} \quad m = 0 \text{ to } 3,$$
$$\text{and} \quad P^\star + Q_o = 1 - J/\Delta.$$

Solving Eq. (A.20) and choosing the root which is well behaved for $x$ near 0, we get

$$Q(x) = \frac{1 - \sqrt{1 - 4(J/\Delta)x(P^\star + Q_0)}}{2(J/\Delta)x} - P^\star, \qquad (A.23)$$

and the expression for $G(x|h)$ [Eq. (A.22)] becomes

$$G(x|h) = \frac{x}{2\Delta}\left[P^\star + Q(x)\right]^3. \qquad (A.24)$$

Expanding $G(x)$ in power series of $x$, we obtain the Eq. (3.19) of sec. 3.2.2.

In the region C, $p_3$ saturates to the value 1, $\phi(-3J-h)$ becomes zero and $(p_3 - p_2)$ is no longer independent of $h$. Substituting the appropriate expressions, we find that

$$Q(x) = \frac{1 - \sqrt{1 - 4(J/\Delta)x[(1 - 3(J/\Delta) - \epsilon) + (P^\star + Q_0)]}}{2(J/\Delta)x} - P^\star, \qquad (A.25)$$

and

$$G(x|h) = \frac{x}{2\Delta}\left\{[P^\star + Q(x)]^3 - P^{\star 3}\right\}. \qquad (A.26)$$

We note that the term inside the radical sign in $Q(x)$, and also in $G(x|h)$, is a simple linear function of $x$. It is thus straightforward to expand it in powers of $x$ using binomial expansion. This gives us the Eq. (3.23) of sec. 3.2.2.

# Bibliography

Adler J. (1991), *Physica A*, **171**, 453.

Aizenman M. and Lebowitz J. L. (1988), *J. Phys. A: Math. Gen*, **21**, 3801.

Aizenmann M. and Wehr J. (1989), *Phys. Rev. Lett.*, **62**, 2503.

Ananikyan N. S., Izmailyan N. S. and Shcherbakov R. R. (1994), *JETP Lett.*, **59**, 71, ; Pis'ma Zh. Eksp. Fiz. **59**, 71 (1994).

Bak P. (1997), *How Nature Works*, Oxford University Press, Oxford.

Barkhausen H. (1919), *Z. Phys.*, **20**, 401.

Bertotti G. (1998), *Hysteresis in Magnetism: for physicists, materials scientists, and engineers*, Academic Press, San Diego.

Bruinsma R. (1984), *Phys. Rev. B*, **30**, 289.

Cerf R. and Cirillo E. N. M. (1999), *Ann. Probab.*, **27**, 1833.

Chakrabarti B. K. and Acharyya M. (1999), *Rev. Mod. Phys*, **71**, 847.

Chalupa J., Leath P. L. and Reich G. R. (1981), *J. Phys. C*, **12**, L31.

Cote P. J. and Meisel L. V. (1991), *Phys. Rev. Lett.*, **67**, 1334.

Dahmen K. and Sethna J. P. (1993), *Phys. Rev. Lett.*, **71**, 3222.

Dahmen K. and Sethna J. P. (1996), *Phys. Rev. B*, **53**, 14872.

Dhar D. (1999), *cond-mat/9909009*.

Dhar D. and Thomas P. B. (1992), *J. Phys. A: Math. Gen*, **25**, 4967.

Dhar D. and Thomas P. B. (1993), *J. Phys. A: Math. Gen*, **26**, 3973.

Dhar D., Shukla P. and Sethna J. P. (1997), *J. Phys. A: Math. Gen*, **30**, 5259.

Feynman R. P., Leighton R. B. and Sands M. (1977), *The Feynman Lectures on Physics*, vol. II, Addison-Wesley.

Imbrie J. Z. (1984), **53**, 1747.

Imry Y. and Ma S. K. (1975), *Phys. Rev. Lett.*, **35**, 1399.

Jensen H. J. (1998), *Self-Organized Criticality*, Cambridge University Press.

Kawasaki K. (1972), Kinetics of Ising models, In C. Domb and M. S. Green, eds., *Phase Transition and Critical Phenomena*, vol. 2, 443, Academic Press, London.

Kittel C. (1949), *Rev. Mod. Phys*, **21**, 541.

Kogut P. M. and Leath P. L. (1981), *J. Phys. C*, **14**, 3187.

Middleton A. A. (1992), *Phys. Rev. Lett.*, **68**, 670.

Nattermann T. (1998), Theory of the random field ising model, In A. P. Young, ed., *Spin Glasses and Random Fields*, 277, World Scientific, preprint cond-mat/9705295.

Parisi G. (1988), *Statistical Field Theory*, 195, Addison-Wesley.

Pázmándi F., Zaránd G. and Zemányi G. T. (1999), *Phys. Rev. Lett.*, **83**, 1034.

Perković O., Dahmen K. A. and Sethna J. P. (1995), *Phys. Rev. Lett.*, **75**, 4528.

Preisach F. (1935), *Z. Phys.*, **94**, 277.

Rao M., Krishnamurthy H. H. and Pandit R. (1990), *Phys. Rev. B*, **42**, 856.

Rayleigh J. S. W. (1887), *Philos. Mag.*, **23**, 225.

Sabhapandit S., Shukla P. and Dhar D. (2000), *J. Stat. Phys.*, **98**, 103.

Sabhapandit S., Dhar D. and Shukla P. (2002), *Phys. Rev. Lett.*, **88**, 197202.

Sethna J. P., Dahmen K. A., Kartha S., Krumhans J. A., Roberts B. W. and Shore J. D. (1993), *Phys. Rev. Lett.*, **70**, 3347.

Shukla P. (2000), *Phys. Rev. E*, **62**, 4725.

Shukla P. (2001), *Phys. Rev. E*, **63**, 029102.

Sipahi L. B. (1994), *J. Appl. Phys.*, **75**, 6978.

Somoza A. M. and Desai R. C. (1993), *Phys. Rev. Lett.*, **70**, 3279.

Spasojević D., Bukvić S., Milošević S. and Stanley H. E. (1996), *Phys. Rev. E*, **54**, 2531.

Stauffer D. and Aharony A. (1992), *Introduction to Percolation Theory*, 26–34, Taylor and Francis, London.

Tadić B. (1996), *Phys. Rev. Lett.*, **77**, 3843.

Urbach J. S., Madison R. C. and Markert J. T. (1995), *Phys. Rev. Lett.*, **75**, 276.

Weiss P. (1907), *J. Phys.*, **6**, 661.

Williams H. J. and Shockley W. (1949), *Phys. Rev.*, **75**, 178.